\documentclass[journal=mamobx,manuscript=article]{achemso}
\setkeys{acs}{articletitle=true}

\SectionNumbersOn

\usepackage[version=3]{mhchem} 
\usepackage[T1]{fontenc} 

\usepackage{color}
\usepackage{gensymb}

\author{Xiaolei Xu}
\affiliation{State Key Laboratory of Polymer Physics and Chemistry, Changchun Institute of Applied Chemistry, Chinese Academy of Sciences, Changchun 130022, P. R. China}

\author{Jack F. Douglas}
\email{jack.douglas@nist.gov}
\affiliation{Materials Science and Engineering Division, National Institute of Standards and Technology, Gaithersburg, Maryland 20899, United States}

\author{Wen-Sheng Xu}
\email{wsxu@ciac.ac.cn}
\affiliation{State Key Laboratory of Polymer Physics and Chemistry, Changchun Institute of Applied Chemistry, Chinese Academy of Sciences, Changchun 130022, P. R. China}
\title{Parallel Emergence of Rigidity and Collective Motion in a Family of Simulated Glass-Forming Polymer Fluids}

\abbreviations{IR,NMR,UV}

\begin{document}

\begin{tocentry}

 \centering
 \includegraphics[height=4.0cm]{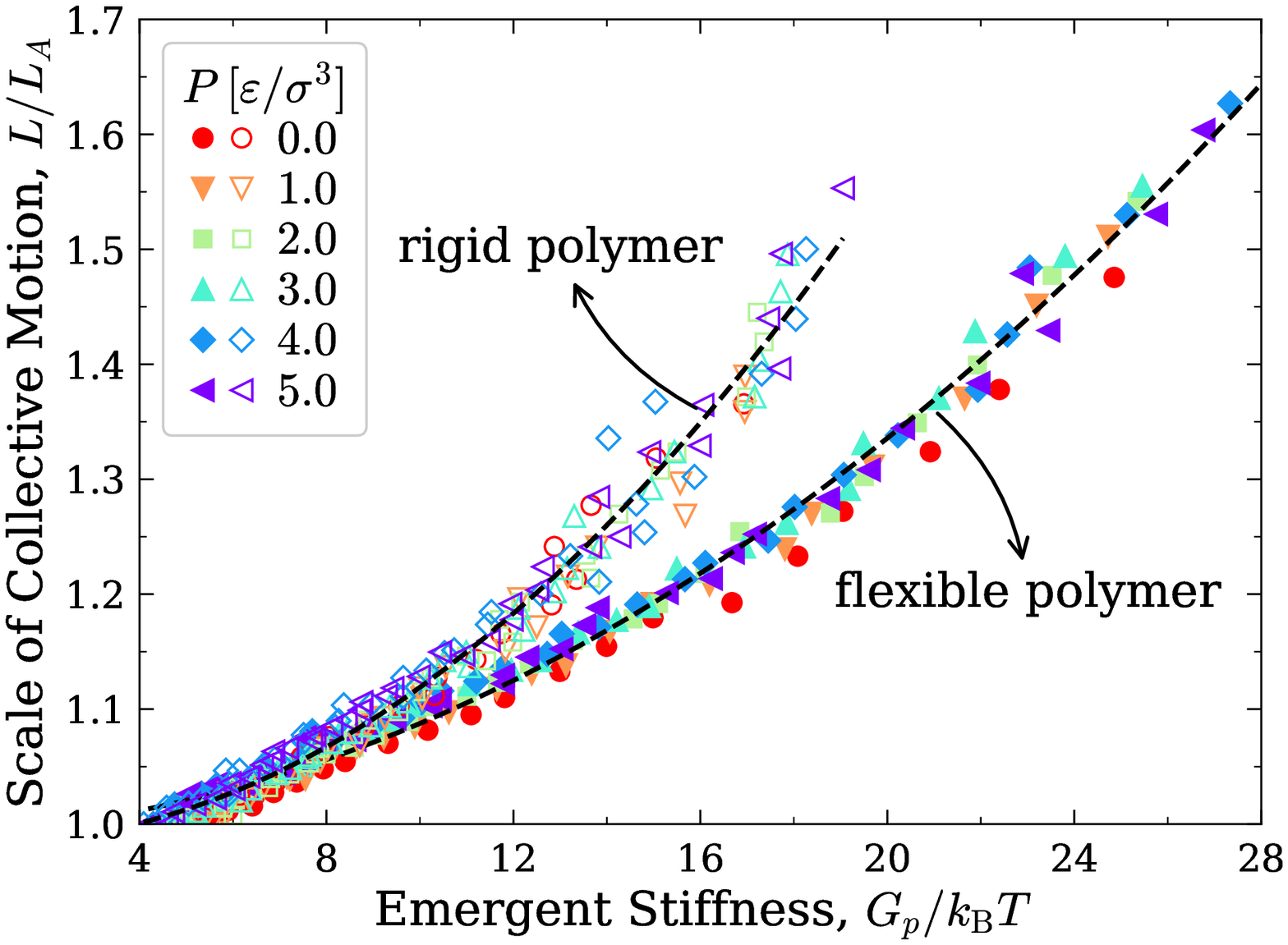}


\end{tocentry}

\newpage

\begin{abstract}
The emergence of the solid state in glass-forming materials upon cooling is accompanied by changes in both thermodynamic and viscoelastic properties and by a precipitous drop in fluidity. Here, we investigate changes in basic elastic properties upon cooling in a family of simulated polymer fluids, as characterized by a number of stiffness measures, such as the ``glassy plateau shear modulus'' $G_p$, the ``non-ergodicity parameter'' $f_{s, q^*}$, the bulk modulus $B$, the Poisson ratio $\nu$, and the ``Debye-Waller parameter'' $\langle u^2 \rangle$, where $G_p$, $f_{s, q^*}$, and $\langle u^2 \rangle$ correspond to the shear stress relaxation function $G(t)$, the self-intermediate scattering function $F_s(q^*, t)$, and the mean square displacement on a ps timescale, respectively. The time dependence of $G(t)$ at elevated temperatures ($T$) resembles the power-law decay predicted by the Rouse model, but stress relaxation transitions to a stretched exponential form in the low $T$ liquid regime dominated by glassy segmental dynamics. In this ``glassy dynamics'' regime, the relaxation times from $G(t)$ and $F_s(q^*, t)$ closely track each other for all polymer models investigated, thereby justifying the identification of the $\alpha$-relaxation time $\tau_{\alpha}$ from $F_s(q^*, t)$ with the structural relaxation time $\tau_{_G}$ from $G(t)$. We show that $\tau_{\alpha}$ can be expressed quantitatively both in terms of measures of the material ``stiffness'', $G_p$ and $\langle u^2 \rangle$, and the extent $L$ of cooperative particle exchange motion in the form of strings, establishing a direct relation between the growth of emergent elasticity and collective motion. Moreover, the macroscopic stiffness parameters, $G_p$, $B$, and $f_{s, q^*}$, can all be expressed quantitatively in terms of the molecular scale stiffness parameter, $k_{\mathrm{B}}T / \langle u^2 \rangle$ with $k_{\mathrm{B}}$ being Boltzmann's constant, and we discuss the thermodynamic scaling of these properties. We also find that $G_p$ is related to the cohesive energy density $\Pi_{\mathrm{CED}}$, pointing to the critical importance of attractive interactions in the elasticity and dynamics of glass-forming liquids. Finally, we discuss fluctuations in the local stiffness parameter as a quantitative measure of elastic heterogeneity and their significance for understanding both the linear and nonlinear elastic properties of glassy materials.
\end{abstract}

\newpage

\section{\label{Sec_Intro}Introduction}

Leading models of relaxation in glass-forming (GF) liquids emphasize the emergence of collective motion~\cite{1965_JCP_43_139, 2013_JCP_138_12A541, 2014_JCP_140_204509, 2021_Mac_54_3001} and material rigidity.~\cite{1981_PRB_24_904, 1980_PRB_22_3130, 2009_PRB_79_132204, 2006_RMP_78_953, 2012_JCP_136_224108, 2015_JNCS_407_14} Both approaches have enjoyed some support in comparison to experiment, leading to the possibility that there are some strong interrelations between them. To directly address this possibility, we consider a family of coarse-grained polymer melts for which the extent $L$ of string-like collective motion and the fragility of glass formation have been extensively studied in the past through changes in the chain stiffness and cohesive energy parameters.~\cite{2016_Mac_49_8341, 2016_Mac_49_8355, 2020_Mac_53_4796, 2020_Mac_53_9678} In the present work, we extend these works to quantify the shear stress relaxation function $G(t)$ and associated properties in the same family of polymer models to enable a quantitative examination of how changes in the extent of collective motion are related to changes in the material rigidity.

After defining the family of polymer models and our simulation methodology, we quantify the self-intermediate scattering function $F_s(q^*, t)$, mean square displacement (MSD) $\langle r^2(t) \rangle$, non-Gaussian parameter $\alpha_2(t)$, etc., as we and others have done in many previous simulation studies,~\cite{2013_JCP_138_12A541, 2014_JCP_140_204509, 2016_Mac_49_8341, 2016_Mac_49_8355, 2016_MacroLett_5_1375, 2017_Mac_50_2585, 2020_Mac_53_4796, 2020_Mac_53_9678} along with essential properties describing elastic and stress relaxation properties of our model GF polymer fluids to allow for a comparative analysis. Here, $q^*$ is a specific wave number and $t$ is the time. As a primary finding, the relaxation times from $F_s(q^*, t)$ and $G(t)$ closely track each other, along with many other parallels between these relaxation functions. We particularly focus on the infinite frequency shear modulus $G_{\infty}$ and the ``glassy plateau shear modulus'' $G_p$ as a function of temperature ($T$), where $G_p$ is often measured as the high frequency shear modulus in viscoelastic materials, since both measures of material stiffness have been suggested to determine the $T$-dependent activation free energy $\Delta G(T)$ of GF liquids.~\cite{1965_JCP_43_139, 2013_JCP_138_12A541, 2014_JCP_140_204509, 2021_Mac_54_3001} We also determine the Debye-Waller parameter $\langle u^2 \rangle$, corresponding to the segmental MSD at the fast $\beta$-relaxation time, since $k_{\mathrm{B}} T / \langle u^2 \rangle$ has been suggested to be an alternative measure of material stiffness,~\cite{2010_SoftMatter_6_292, 2015_EPJE_38_87, 2021_JCP_155_204504, 2022_JCP_157_064901, 2022_Mac_55_9990} where $k_{\mathrm{B}}$ is Boltzmann's constant. In particular, an approximate linear relation between $G_p$ and $k_{\mathrm{B}} T / \langle u^2 \rangle$ has been found to hold for a wide range of materials,~\cite{2015_EPJE_38_87, 2021_JCP_155_204504, 2022_JCP_157_064901, 2022_Mac_55_9990} and we confirm this relation in the present work for our family of coarse-grained polymer models. Moreover, we show that the structural relaxation time $\tau_{\alpha}$ from $F_s(q^*, t)$ can be expressed quantitatively both in terms of measures of the material ``stiffness'', $G_p$ and $\langle u^2 \rangle$, supporting a direct relation between the growth of emergent elasticity and the relaxation time. The glassy plateau shear modulus $G_p$, the bulk modulus $B$, and the non-ergodicity parameter $f_{s, q^*}$ defining the transient plateau in $F_s(q^*, t)$ can all be expressed quantitatively in terms of the molecular scale stiffness parameter, $k_{\mathrm{B}} T / \langle u^2 \rangle$. We also quantify the emergence of collective motion in our family of polymer models through a fit of our simulation data to the string model of glass formation.~\cite{2013_JCP_138_12A541, 2014_JCP_140_204509} This model also describes our relaxation data very well and provides a quantitative measure of collective motion that is in broad accord with the heuristic idea of the cooperatively rearranging regions of Adam and Gibbs (AG),~\cite{1965_JCP_43_139} in the sense that the mass of the string-like clusters determines the change of the activation free energy for relaxation. The simulation observations then puts us in a position to directly examine the interrelation between the growing elasticity and collective motion in GF liquids upon cooling towards the glass transition temperature ($T_{\mathrm{g}}$).

Our analysis reveals that $G_p$ indeed grows in a parallel fashion to $L$ so that the material stiffness grows in lockstep with the extent of collective motion in these models of GF polymer fluids. Both of these properties correlate rather well with the growth of the activation free energy $\Delta G(T)$ of $\tau_{\alpha}$. This relationship is highly quantitative and general in the case of $L$ and $\Delta G(T)$. We also show that the linear relation between $G_p$ and $k_{\mathrm{B}} T / \langle u^2 \rangle$ holds to a good approximation for our model polymers so that $\Delta G(T)$ can be described rather well in terms of $\langle u^2 \rangle$. Our results then indicate that emergent collective motion and rigidity represent two sides of the same phenomenon of the slowing down of dynamics of GF liquids. A previous work, based on a similar coarse-grained polymer model, has shown that $L$ also scales inversely with the configurational entropy $S_c$.~\cite{2013_JCP_138_12A541} Hence, the emergent rigidity and collective motion also occur hand in hand with the reduction of $S_c$, as anticipated in the AG theory,~\cite{1965_JCP_43_139} and the more recent generalized entropy theory (GET)~\cite{2021_Mac_54_3001} and string model~\cite{2014_JCP_140_204509} of glass formation, which grew out of the pioneering AG model. We also note that the quantity $\langle u^2 \rangle^{3/2}$ can be viewed as a variant of \textit{dynamical free volume},~\cite{1972_JCP_57_1259, 1979_JCP_70_1837, 1986_JPC_90_6252, 1998_MP_95_289, 1997_PRE_56_5524, 2012_SoftMatter_8_11455, 2021_Mac_54_3247, 2022_Mac_55_8699} corresponding to the average volume explored by a particle within its cage arising from the presence of surrounding particles. This interpretation of $\langle u^2 \rangle$ forms the basis of the localization model of glass formation.~\cite{2012_SoftMatter_8_11455, 2015_PNAS_112_2966, 2016_JSM_054048} We thus obtain some unity in our understanding of the relation between leading models of glass formation. Evidently, there are multiple equally valid perspectives of the dynamics of GF liquids, each of which provides insights into particular aspects of glass formation.

One of the phenomenological characteristics of GF liquids that can serve to test the consistency with corresponding theories of glass formation is the property of ``thermodynamic scaling''.~\cite{2005_RPP_68_1405, 2010_Mac_43_7875, Book_Roland, Book_Paluch, 1971_JCP_55_1128} Specifically, the structural relaxation time and other dynamical properties can be expressed in terms of a ``universal'' reduced variable, $TV^{\gamma_t}$, where $\gamma_t$ is a scaling exponent describing how the temperature $T$ and the volume $V$ are linked to each other when either quantity is varied. The existence of the exponent $\gamma_t$ describing the reduced scaling of both thermodynamic and dynamical properties can readily be deduced for idealized materials at constant volume whose particles interact through decomposable pairwise repulsive interactions.~\cite{2009_JCP_131_234504, 2019_JCP_151_204502} This scaling property can be extended to fluids having relatively ``simple'' pairwise potentials,~\cite{2009_JCP_131_234504, 2019_JCP_151_204502} but the origin of scaling property is currently not clear in many molecular fluids where there are additional bonding and rotational potentials, and often long-range interactions due to dipolar and charged species.~\cite{2021_Mac_54_3247, 2022_Mac_55_8699} Nevertheless, this scaling property seems to be broadly observed experimentally to a good approximation. It then becomes a matter of interest to determine whether this scaling property applies to the dynamical properties of our model polymer fluids and various thermodynamic properties proposed to model the dynamics of GF liquids. We find that $\tau_{\alpha}$ obeys thermodynamic scaling to an excellent approximation, as found in many previous experimental and simulation studies.~\cite{2005_RPP_68_1405, 2010_Mac_43_7875, Book_Roland, Book_Paluch} Moreover, $L$, $G_p / k_{\mathrm{B}} T$, and $\langle u^2 \rangle$ also share this scaling property to a good approximation. On the other hand, other properties, such as the low angle scattering intensity $S(0)$ and $G_{\infty} / k_{\mathrm{B}} T$, do not exhibit thermodynamic scaling, where $S(0)$ is related to the isothermal compressibility $\kappa_T$, density $\rho$, and $T$ through the definition, $S(0) = \rho k_{\mathrm{B}} T \kappa_T$, appropriate in general for materials in equilibrium. We note that $S(0)$ is a primary measure of how fluid ``structure'' relates to its thermodynamic state and figures prominently in some models of glass formation, such as the mode-coupling theory.~\cite{Book_Gotze, 2015_PRL_115_205701, 2016_JSM_2016_054049, 2018_FP_6_97, 2022_PRL_129_145501} We make some tentative suggestions for why some thermodynamic properties exhibit thermodynamic scaling, while this scaling property does not hold for other properties that might be expected to exhibit this property based on an assumed effective power-law intermolecular interaction potential. Thermodynamic scaling then offers some hints about the extraordinary interrelationships between the thermodynamics and dynamics of GF liquids.~\cite{2021_Mac_54_3247}

In our recent work,~\cite{2022_Mac_55_8699} we investigated linkages between the thermodynamic properties of polymeric melts, and provided evidence supporting a free energy landscape origin of relationships between thermodynamic and dynamic properties of GF liquids. Relationships of this kind are assumed in the AG model,~\cite{1965_JCP_43_139} and this type of assumption forms the basis of many models of the dynamics of GF liquids, although the assumptions on which these models are based are frankly hard to justify. As a continuation of this general theoretical theme from a somewhat different angle, we show that fundamental elasticity measures $G_p$ and $\langle u^2 \rangle$ are highly correlated with the cohesive energy density of the fluid, $\Pi_{\mathrm{CED}}$. These relationships not only illuminate the origin of thermodynamic scaling for some selected thermodynamic properties related to $\Pi_{\mathrm{CED}}$, but also serve to emphasize the critical importance of attractive intermolecular interactions for understanding both the elastic and relaxation properties of GF liquids, given the relation of $G_p$ and $\langle u^2 \rangle$ to these properties. We also expect these relations to be highly useful in practical material design, because an existing knowledge of the how molecular structure and thermodynamic conditions influence the cohesive energy density of polymer materials is based on numerous former experimental studies and helpful theoretical models of $\Pi_{\mathrm{CED}}$. We further extend our analysis of the characterization of the elastic properties of our family of coarse-grained polymers to describe elasticity heterogeneity on a nanoscale. In particular, we apply the relation between $G_p$ and $\langle u^2 \rangle$, established as applying to these materials at a macroscale, to make a map of the stiffness fluctuations at a nanoscale. The resulting ``stiffness field'' map enables the facile visualization of stiffness fluctuations of the polymer material. Based on this map, we conspicuously observe regions of the material in which chain segments are strongly localized and immobile, and also complementary regions in which the local packing tends to be frustrated and where the polymer segments are relatively delocalized and much more mobile. It is in these ``softer'' regions, reminiscent of grain boundaries in polycrystalline materials,~\cite{2009_PNAS_106_7735, 2010_PRE_81_041301, 2011_PNAS_108_11323} where the particle motion in the form of strings is normally concentrated. A similar view of the dynamics of cooled liquids in terms of well-packed and immobile regions surrounded by an interstitial ``connective tissue'' of relatively loosely packed and more mobile molecules was suggested by Johari~\cite{2002_JNCS_307_310_317} in his model of the ``Johari-Goldstein (JG) relaxation process'', an apparently general relaxation process in GF liquids that occurs after the fast $\beta$-relaxation process on a ps timescale, but before the $\alpha$-relaxation process. Accordingly, the Johari-Goldstein relaxation process is often said to be a ``precursor'' of the $\alpha$-relaxation process. In particular, Johari hypothesized that this type of relaxation occurred within ``islands of mobility'' within a ``connective tissue'' of relatively low mobility. We show below through a direct imaging of regions of relative high and low mobility in our simulated polymeric GF materials that Johari's physical perspective of GF liquids is supported by our simulation observations.

In addition to providing many qualitative insights into the macroscopically averaged properties, such as the origin of decoupling in terms of the asymmetry in the lifetimes of the immobile and mobile particle domains, the visualization of elasticity heterogeneity seems to offer great potential for understanding some of the unique nonlinear deformation properties of polymer materials in their glass state. We provide some illustrations of the ``stiffness maps'' for polymers having different molecular parameters and under conditions of variable pressure ($P$) to rationalize certain widely observed patterns of behavior in the plasticity of glassy polymer materials. These stiffness maps should also be of interest to compare with recent observations of nanoscale stiffness fluctuations in both polymeric and metallic GF materials.

\section{Model and Simulation Details}

\subsection{\label{Sec_Model}Model}

Our investigation of polymer glass formation is based on a generic, coarse-grained model of polymers.~\cite{1990_JCP_92_5057, 1986_PRA_33_3628} This model describes the basic characteristics of polymers by representing the chains in terms of a certain number of connected statistical segments, which are commonly called ``beads''. A similar model has been widely used for studying glass formation.~\cite{2020_Mac_53_4796, 2020_Mac_53_9678, 2020_Mac_53_6828, 2021_Mac_54_3247, 2022_Mac_55_8699, 2021_Mac_54_6327, 2022_Mac_55_3221}

Our coarse-grained polymer model contains both bonded and nonbonded interactions. First, to maintain chain connectivity, neighboring beads along a chain are bonded to each other through the finitely extensible nonlinear elastic (FENE) potential,~\cite{1990_JCP_92_5057, 1986_PRA_33_3628}
\begin{eqnarray}
	\label{Eq_FENE}
	U_{\mathrm{FENE}}(r) = -\frac{1}{2} k_b R_0^2 \ln\left[1 - \left(\frac{r}{R_0}\right)^2\right] + 4 \varepsilon \left[ \left(\frac{\displaystyle \sigma}{\displaystyle r}\right)^{12} - \left(\frac{\displaystyle \sigma}{\displaystyle r}\right)^6 \right] + \varepsilon
\end{eqnarray}
where $r$ denotes the distance between two beads and $\varepsilon$ and $\sigma$ are the energy and length scales associated with the Lennard-Jones (LJ) potential. The first term of eq~\ref{Eq_FENE} extends to $R_0$, with the common choices of $k_b = 30 \varepsilon/\sigma^2$ and $R_0 = 1.5 \sigma$, and the second term has a cutoff at $2^{1/6} \sigma$. Moreover, chain rigidity can be controlled by applying an angular potential to two consecutive bonds,~\cite{2019_JCP_150_091101}
\begin{eqnarray}
	\label{Eq_Bend}
	U_{\mathrm{bend}}(\theta) = -A \sin^2(B \theta),\ 0 < \theta < \pi / B
\end{eqnarray}
where the bond angle is given by $\theta = \cos^{-1}[(\mathbf{b}_{j} \cdot \mathbf{b}_{j+1})/ (|\mathbf{b}_{j}| |\mathbf{b}_{j+1}|)]$ in terms of the bond vector $\mathbf{b}_{j} = \mathbf{r}_{j} - \mathbf{r}_{j-1}$ between two neighboring beads $j$ and $j-1$. The parameter associated with the rest angle is fixed at $B = 1.5$ based on our previous work,~\cite{2020_Mac_53_4796} so a larger value of $A$ results in a stiffer chain. Finally, the following truncated-and-shifted LJ potential,
\begin{eqnarray}
	\label{Eq_LJ}
	U_{\mathrm{LJ}}(r) = 4 \varepsilon \left[ \left(\frac{\displaystyle \sigma}{\displaystyle r}\right)^{12} - \left(\frac{\displaystyle \sigma}{\displaystyle r}\right)^6 \right] + C(r_{\mathrm{cut}}),\ r < r_{\mathrm{cut}}
\end{eqnarray}
is used to model the nonbonded interactions between any two nonbonded beads. Here, $C(r_{\mathrm{cut}})$ is a constant to ensure that $U_{\mathrm{LJ}}$ varies smoothly to zero at the cutoff distance, $r_{\mathrm{cut}} = 2.5\sigma$, a choice that is intended to include attractive nonbonded interactions.

To check the generality of our findings, we investigate the glass formation of a range of polymer melts having variable chain length $M$ and chain rigidity $A$. Glass formation is also studied under different constant $P$ conditions so that we can utilize thermodynamic scaling to gain further insight into the rigidity and collective motion of GF liquids. The total number of beads is $N = 12000$ in most of our simulations, except for the polymer melt with $M = 20$ and $A = 0 \varepsilon$ at varying $P$, where a system of $N = 8000$ is adopted instead. Our results indicate that the mass effects on the characteristic temperatures of glass formation saturate after a moderate chain length in the present coarse-grained polymer model. In particular, the chain length of $20$ beads already lies in a mass range where the segmental properties are relatively insensitive to changes in $M$. We thus expect that the main results and conclusions remain unchanged if longer chains are studied.

We describe all quantities and results in standard reduced LJ units. Specifically, length, time, temperature, and pressure are, respectively, given in units of $\sigma$, $\tau$, $\varepsilon / k_{\mathrm{B}}$, and $\varepsilon / \sigma^3$, where $\tau = \sqrt{m_b \sigma^2 / \varepsilon}$ with $m_b$ being the bead mass. These reduced units may be roughly mapped to laboratory units, e.g., by taking the suggested choices of Baschnagel and coworkers,~\cite{Baschnagel_Chapter} i.e., $\sigma \approx 5 \times 10^{-10}$ m, $\varepsilon / k_{\mathrm{B}} \approx 450$ K, and $m_b \approx 60$ g/mol. A reduced time of $t = 1 \tau$ and a reduced pressure of $P = 1 \varepsilon / \sigma^3$ then correspond approximately to $1$ ps and $50$ MPa, respectively. As a reference, the entanglement length is about $M = 84$ in a similar coarse-grained linear polymer melt without bending constraints at a number density of $\rho = N/V = 0.85 \sigma^{-3}$ and a temperature of $T = 1.0 \varepsilon/k_{\mathrm{B}}$.~\cite{2016_JCP_145_141101} The chain lengths considered in the present work are well below the entanglement length. Notably, the bead is generally not equivalent to the Kuhn segment in this model, and the Kuhn length $l_K$ must be measured in simulations.~\cite{2020_Mac_53_1901} For more information about the relationship between the simulation model and real polymers, we refer the reader to ref~\citenum{2020_Mac_53_1901} where mapping relations between the reduced and laboratory units have been provided for a wide range of commodity polymers. These mapping relations are achieved by directly tuning the chain rigidity of the simulation model such that the number of Kuhn segments within the volume of a Kuhn length cube for the simulation model matches that for the target polymer.

Let us briefly comment on the limitations of the coarse-grained polymer modeling. The dynamics in the sub-picosecond regime are due to bond vibrations, angle librations, and local motion of side groups. Such modes arise in the bead-spring model, depending on the level of coarse-graining involved in the modeling. Our coarse-grained model thus does not describe the sub-picosecond dynamics of real polymer melts in a molecularly faithful way. If this is one’s goal, then it is probably necessary to perform first principles quantum calculations rather than classical molecular dynamics simulations. It is not clear, however, that an accurate description of the sub-picosecond dynamics is required to model qualitative trends in the dynamics of GF liquids as a function of molecular parameters and variable thermodynamic conditions such as $P$ and $T$. We believe that coarse-grained polymer models are perfectly suited for this general purpose. Nonetheless, it should be pointed out at the outset that physical phenomena smaller than a polymer segment and processes occurring on a timescale shorter than a ps cannot be reliably described by the coarse-grained model employed in the present paper. We refer the reader to refs~\citenum{2017_Mac_50_4524} and~\citenum{2021_Mac_54_2740} for a detailed discussion of the limitations of the coarse-grained polymer modeling. A comprehensive review on simulation studies of polymer glass formation has been provided by Baschnagel and coworkers.~\cite{2010_SoftMatter_6_3430}

\subsection{\label{Sec_Simulation}Simulation Details}

All our simulations are carried out in three dimensions under periodic boundary conditions utilizing the Large-scale Atomic/Molecular Massively Parallel Simulator (LAMMPS) molecular dynamics package.~\cite{LAMMPS_1995, LAMMPS_webpage} For each polymer system, we initially prepare an equilibrated melt system in the $NPT$ ensemble, where $P$ and $T$ are maintained by a Nos\'{e}-Hoover barostat and thermostat as implemented in LAMMPS and a time step of $\Delta t = 0.005 \tau$ is used to integrate the equations of motion. The temperature is chosen to be sufficiently high so that the polymer melt can be equilibrated properly within our time window. This step allows us to obtain the number density, as determined from a production run of $10^5 \tau$ under the specified thermodynamic conditions after an equilibration of $10^5 \tau$. Following these simulations, the melt with the desired density is further equilibrated for a period of $2 \times 10^5 \tau$ in the $NVT$ ensemble, which is much longer than the longest relaxation time of the melt to ensure the proper equilibration of the polymer melt. The equilibrated melt is subsequently subjected to a cooling or heating process at a constant $P$ at a rate of $10^{-4} \varepsilon / (k_{\mathrm{B}} \tau)$, which enables us to obtain the density as a function of $T$. In our simulations, we focus on a $T$ regime well above $T_{\mathrm{g}}$ so that our results are not complicated by the nonequilibirum effects associated with the glass state. Nonequilibrium effects associated with the propensity for the liquid to crystallize found in many atomic liquids and mixtures of small-molecule liquids are avoided because our polymer fluids do not appear to be capable of crystallizing so that we do not refer to our polymer fluids at low $T$ as being ``supercooled''. Properties are calculated in the $NVT$ ensemble after the melt is further equilibrated for a period typically over $10$ to $100$ times longer than $\tau_{\alpha}$. We admit that there is no actual way to prove the material in true thermodynamic equilibrium in the precise sense prescribed by Gibbs and Boltzmann.

\section{Results and Discussion}

\subsection{\label{Sec_Definition}Basic Properties of Glass-Forming Polymer Fluids}

We begin with an introduction to the definitions of basic properties of glass formation considered in the present paper. The segmental structural relaxation time $\tau_{\alpha}$ is one of the most fundamental properties characterizing the dynamics of GF polymer fluids. The determination of $\tau_{\alpha}$ is based on the self-part of the intermediate scattering function,
\begin{eqnarray}
	F_s(q, t) = \frac{1}{N}\left< \sum_{j=1}^N\exp\{-i \mathbf{q} \cdot [\mathbf{r}_j(t) - \mathbf{r}_j(0)]\} \right>
\end{eqnarray}
where $i$ is the imaginary unit, $q = |\mathbf{q}|$ is the wave number, $\mathbf{r}_j$ is the position of particle $j$, $t$ is the time, and the angular brackets $\langle \cdot\cdot\cdot \rangle$ denote the usual thermal average. The wave number is chosen to be $q = q^* = 7.0 \sigma^{-1}$, which is close to the first peak position of the static structure factor $S(q)$. Following previous works,~\cite{2013_JCP_138_12A541, 2020_Mac_53_4796} we define $\tau_{\alpha}$ by the time where $F_s(q^*, t)$ decays to a value of $0.2$. Alternatively, $\tau_{\alpha}$ might be determined by fitting $F_s(q^*, t)$ to a stretched exponential function.~\cite{2020_SciAdv_6_eaaz0777} $F_s(q^*, t)$ also enables the definition of a fixed time ``non-ergodicity parameter'' at short $t$, $f_{s, q^*} = F_s(q^*, t = 1 \tau)$.

The determination of the average string length $L$ follows the previously established method for polymeric systems.~\cite{2003_JCP_119_5920, 2013_JCP_138_12A541} Mobile particles are defined as the $f_0=6.5\%$ of particles with the greatest displacement over any chosen interval. Two mobile particles $j$ and $k$ are then considered to be in the same string if
\begin{eqnarray}
	\min [|\mathbf{r}_j(t) - \mathbf{r}_k(0)|, |\mathbf{r}_k(t) - \mathbf{r}_j(0)|] < \delta
\end{eqnarray}
where $\delta$ is a displacement distance parameter that the present work chooses to be $0.55$. The number averaged string length $\langle s(t) \rangle$ is then calculated via $\langle s(t) \rangle = \langle \sum_{s=1}^{\infty}sC(s) \rangle/\sum_{s=1}^{\infty}C(s)$,
with $C(s)$ being the probability of finding a string of length $s$. The quantity $\langle s(t) \rangle$ is found to display a maximum $L \equiv \langle s(t_L) \rangle$ at a characteristic time $t_L$, where $L$ defines the characteristic string length defining the ``extent'' of collective motion.

The Debye-Waller parameter $\langle u^2 \rangle$ is determined from the segmental MSD at a characteristic time on the order of the decay time of the intermediate scattering function typically on the order of $1$ ps in molecular liquids,~\cite{2018_JCP_148_104508}
\begin{eqnarray}
	\langle r^2(t) \rangle = \frac{1}{N} \left\langle \sum_{j=1}^N |\mathbf{r}_j(t) - \mathbf{r}_j(0)|^2 \right\rangle
\end{eqnarray}
In particular, $\langle u^2 \rangle$ is defined by the value of $\langle r^2(t) \rangle $ at the ``caging time'' of $t = 1 \tau$ in our dimensionless time units. Notably, the MSD allows us to obtain the onset temperature $T_A$ of glass formation,~\cite{2016_Mac_49_8341, 2016_Mac_49_8355, 2020_Mac_53_4796, 2020_Mac_53_9678} at which the dynamics starts to exhibit non-Arrhenius behavior characteristic of GF materials. $T_A$ is also called the ``localization temperature'' since it is clearly related to the particle localization on a ps timescale. As a complement to the MSD, we also consider the non-Gaussian parameter,
\begin{eqnarray}
	\alpha_2(t) = \frac{3 \langle r^4(t) \rangle}{5 \langle r^2(t) \rangle^2} - 1
\end{eqnarray}
This property provides a characterization of the ``dynamic heterogeneity'' associated with the mobile particles, which exhibit collective motion and dominate molecular diffusion processes that enable structural relaxation. A primary peak $\alpha_2^*$ is generally exhibited in $\alpha_2(t)$ at a peak time $t^*$,~\cite{2018_JCP_149_234904} which is related to the lifetime of the mobile particle clusters.~\cite{2013_JCP_138_12A541, 2015_JCP_142_164506} Alternatively, $t_L$ and $L$ can be used to characterize the spatially heterogeneous motion of the mobile particles. Despite the correlation with the mobile particle clusters just mentioned, the physical significance of $\alpha_2(t)$ is somewhat uncertain. Song et al.~\cite{2019_PNAS_116_12733} and Zhang and Douglas~\cite{2013_SoftMatter_9_1254} have argued that $\alpha_2(t)$ can be related to a four-point velocity correlation function, which is a natural measure of mobility fluctuations. Further work is needed to better elucidate the physical meaning of this widely studied property of polymeric and other complex fluids.

To characterize the dynamic heterogeneity from the perspective of immobile particles, we study the self-part of the four-point susceptibility or ``caged'' particle clusters responsible for the growth of the structural relaxation time upon cooling,~\cite{2003_JCP_119_7372} 
\begin{equation}
	\chi_{4, s}(t) = \frac{V}{N^2}\left[ \langle Q_{s}(t)^{2} \rangle - \langle Q_{s}(t) \rangle^{2} \right]
\end{equation}
Here, the self-overlap order parameter is defined by $Q_{s}(t) = \sum_{j=1}^{N} w(|\textbf{r}_{j}(t) - \textbf{r}_{j}(0)|)$, where $w = 1$ for $|\textbf{r}_{j}(t) - \textbf{r}_{j}(0)| < 0.3 \sigma$ and zero otherwise. Notice that $\chi_{4, s}(t)$ is also sensitive to the magnitude of the cutoff, as in the calculation of $\langle s(t) \rangle$. A recent work has discussed the precise criterion of this cutoff in liquids generally.~\cite{2019_JCP_151_184503} With the cutoff taken to have the assumed value just found for the binary LJ liquid, $\chi_{4, s}(t)$ exhibits a peak $\chi_{4,s}^{*}$ at an intermediate time $t_{\chi}$ that defines the lifetime of the immobile particles, where both $\chi_{4,s}^{*}$ and $t_{\chi}$ grow upon cooling. 

The novel part of the present work primarily pertains to properties in relation to the stress autocorrelation function,
\begin{equation}
	\label{Eq_GT}
	G(t) = \frac{V}{k_{\mathrm{B}} T} \left< \sigma_{xy}(t)\sigma_{xy}(0) \right>
\end{equation}
where $\sigma_{xy}$ is the off-diagonal component of the stress tensor. To calculate this property, we have utilized the multiple-tau correlator method of Ram\'{i}rez et al.~\cite{2010_JCP_133_154103} The glassy plateau shear modulus $G_p$, introduced before by Leporini and coworkers,~\cite{2012_JCP_136_041104, 2016_JCP_145_234904, 2017_JPCM_29_135101} is estimated from $G(t)$ at the ``caging time'' $t = 1 \tau$, i.e., $G_p = G(t = 1 \tau)$, as in the case of $\langle u^2 \rangle$, while $G_{\infty}$ is determined from the limit of $G_{\infty} = G(t \rightarrow 0_{+})$. $G(t)$ is an extremely important property for understanding glass formation, but this property has received limited attention from a computational perspective.

\begin{figure*}[htb]
	\centering
	\includegraphics[angle=0,width=0.8\textwidth]{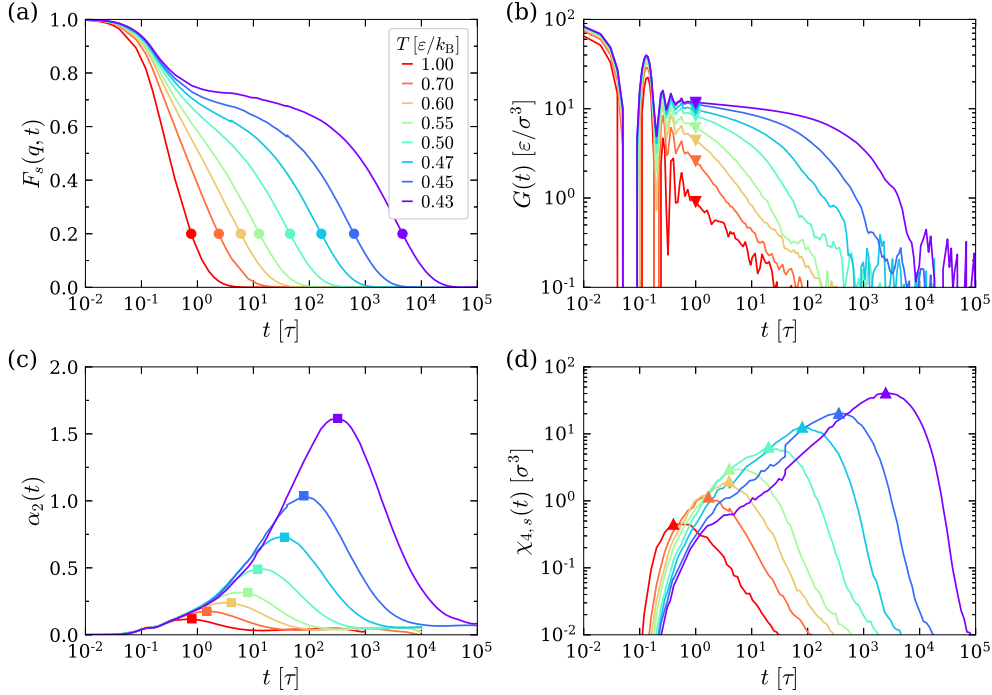}
	\caption{\label{Fig_TCF}Basic properties of glass formation considered in the present work. Panels (a--d) show the self-intermediate scattering function $F_s(q, t)$, the stress autocorrelation function $G(t)$, the non-Gaussian parameter $\alpha_2(t)$, and the four-point susceptibility $\chi_{4, s}(t)$, respectively, as a function of time $t$ at varying temperatures $T$ for the polymer melt having the chain length of $M = 20$ and the chain rigidity parameter of $A = 0 \varepsilon$ at a pressure of $P = 0.0 \varepsilon / \sigma^3$. The wave number is chosen to be $q = q^* = 7.0 \sigma^{-1}$ for the calculation of $F_s(q, t)$. The definitions of the structural relaxation time $\tau_{\alpha}$, the glassy plateau shear modulus $G_p$, the peak magnitude of $\alpha_2(t)$, and the peak magnitude of $\chi_{4,s}(t)$ are indicated by the symbols.}
\end{figure*}

As an illustration, Figure~\ref{Fig_TCF} shows the $t$ dependences of $F_s(q^*, t)$, $G(t)$, $\alpha_2(t)$, and $\chi_{4,s}(t)$ at varying $T$ for the polymer melt having the chain length of $M = 20$ and the chain rigidity parameter of $A = 0 \varepsilon$ at $P = 0.0 \varepsilon / \sigma^3$. The emphasis here is that all the properties considered exhibit significant changes upon cooling. While our previous simulation works~\cite{2016_Mac_49_8341, 2016_Mac_49_8355, 2016_MacroLett_5_1375, 2017_Mac_50_2585, 2020_Mac_53_4796, 2020_Mac_53_6828} have discussed the properties associated with some of the time correlation functions described above, such as $F_s(q^*, t)$, $\alpha_2(t)$, and $\chi_{4,s}(t)$, the primary focus of the present work is on the properties associated with $G(t)$ and their connections to other characteristic properties of glass formation.

We comment on the origin of the large oscillations of $G(t)$ at short $t$ in Figure~\ref{Fig_TCF}b. As discussed by Likhtman and coworkers,~\cite{2007_Mac_40_6748} this behavior arises due to the bond length relaxation. This has also been noted in a recent work by Baschnagel and coworkers,~\cite{2019_JCP_151_054504} who utilizes a similar polymer model with the bond potential being of harmonic form. Puosi and Leporini~\cite{2012_JCP_136_041104} have shown that it is possible to remove these fast oscillations by performing averages over short time intervals to obtain a ``coarse-grained'' dynamics. This additional averaging procedure is helpful in visualizing the relaxation process of $G(t)$, but we avoid this procedure in the present work. Notably, our results of $G(t)$ are fully in line with state of the art calculations by Baschnagel and coworkers~\cite{2017_PRL_119_147802, 2018_PRE_97_012502, 2019_JCP_151_054504} and Leporini and coworkers,~\cite{2012_JCP_136_041104, 2015_JPSB_53_1401} which are based on similar coarse-grained polymer models.

\begin{figure*}[htb]
	\centering
	\includegraphics[angle=0,width=0.8\textwidth]{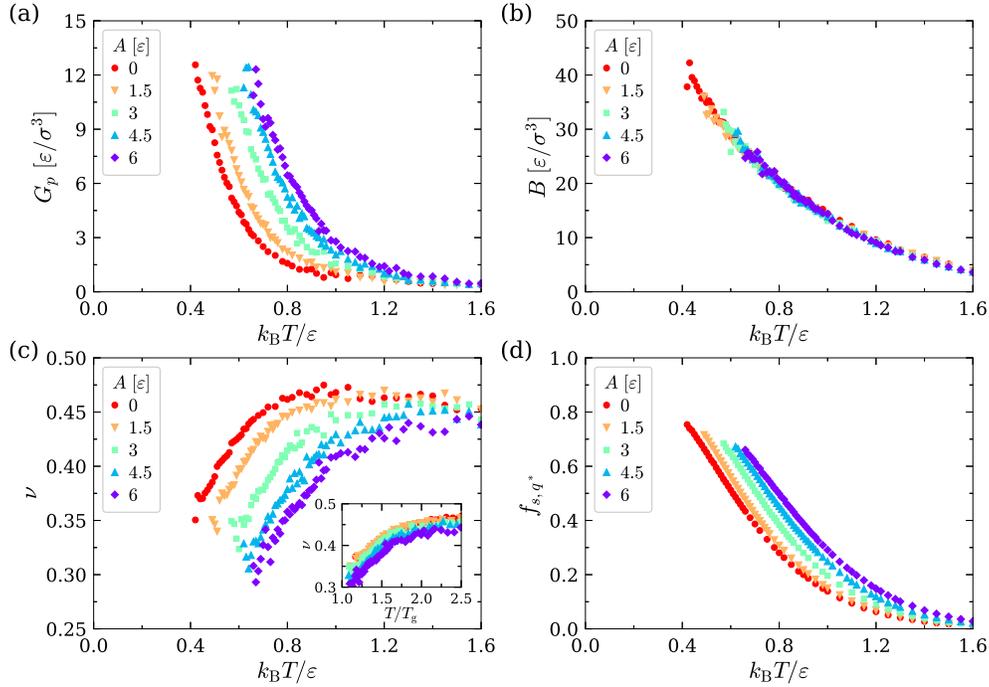}
	\caption{\label{Fig_GpT}Temperature dependence of various properties of glass formation. Panels (a--d) show the glassy plateau shear modulus $G_p$, the bulk modulus $B$, the Poisson ratio $\nu$, and the non-ergodicity parameter $f_{s, q^*}$ as a function of $T$ for polymer melts having variable $A$ at $P = 0.0 \varepsilon / \sigma^3$, respectively. The inset in panel (c) shows $\nu$ as a function of $T / T_{\mathrm{g}}$. The chain length is fixed at $M = 20$.}
\end{figure*}

To illustrate the $T$ variation of $G_p$, we show $G_p$ as a function of $T$ for polymer melts having variable chain rigidity $A$ at $P = 0.0 \varepsilon / \sigma^3$ in Figure~\ref{Fig_GpT}, where we also indicate the corresponding results of the bulk modulus $B$, Poisson ratio $\nu$, and non-ergodicity parameter $f_{s, q^*}$. Here, the bulk modulus is estimated as $B = 1 / \kappa_T$, where the isothermal compressibility is determined as $\kappa_T = S(0) / (\rho k_{\mathrm{B}}T)$ with $S(0)$ being the long-wavelength limit of $S(q)$, and the Poisson ratio is defined by $\nu = (3B - 2G_p) / (6B + 2G_p)$. Parenthetically, our simulation estimates of $G_p$ are on the order of $10^8$ Pa over the range of $T$ investigated by following the mapping relation between the reduced and laboratory units discussed in Section~\ref{Sec_Model}. These magnitudes are comparable to those of real polymer materials (e.g., see the experimental data in ref~\citenum{2013_MacroLett_2_970}).

We note the remarkable insensitivity of $B$ to the variation of chain stiffness, which is in great contrast to $G_p$ and $f_{s, q^*}$. A previous work has shown that $B$ is also insensitive to molecular mass,~\cite{2020_JCP_153_054902} and we have confirmed that this is the case in our family of polymer models. Figure~\ref{Fig_GpT} further indicates that increasing the chain rigidity, which creates packing frustration~\cite{2008_Mac_41_7232} and thus an increase in the fragility of glass formation,~\cite{2020_Mac_53_4796} leads to an increase in $G_p$ at low $T$. This effect increases progressively upon approaching $T_{\mathrm{g}}$. The combined effect of the relative constancy of $B$ and the increase in $G_p$ as the chains are made stiffer leads to a corresponding decrease in the relative value of the Poission ratio, $\nu$. It is emphasized that our results are limited to the equilibrium fluid regime and that different trends might arise, depending on the thermal history, in materials quenched into a non-equilibrium glass state.~\cite{2010_SoftMatter_6_292} Increasing the stiffness of our polymers evidently leads to a progressive increase in the fragility of glass formation~\cite{2020_Mac_53_4796, 2020_Mac_53_9678} so that we might expect $\nu$ to be anti-correlated in its variation with fragility. This type of correlation, which has been of much interest in the glass science community, accords with the observation of some materials, but Johari~\cite{2002_JNCS_307_310_317, 2006_Nature_442_E7} has emphasized that this type of correlation does not hold in network-forming and associating fluids so that this relation is not universal for all types of GF materials. A recent exhaustive examination of the Poisson ratio in diverse inorganic GF materials~\cite{2019_Materials_12_2439} has confirmed Johari's arguments and observations.

Zheng et al.~\cite{2022_JCP_157_064901, 2022_Mac_55_9990} have extensively discussed the $T$ dependence of $\nu$ in their recent works, so we do not reproduce this discussion here. We point out, however, that $\nu$ does not quantitatively reduce to a universal master curve as a function of $T / T_{\mathrm{g}}$ (see the inset of Figure~\ref{Fig_GpT}c), as observed for a family of thermoset polymer materials having variable cross-link density and cohesive interaction strength.~\cite{2022_JCP_157_064901, 2022_Mac_55_9990} This reduction is apparently not universal, but seems to apply to restricted classes of GF materials.

\subsection{\label{Sec_Decoupling}Decoupling Relation between Different Characteristic Timescales of Glass Formation}

\begin{figure*}[htb]
	\centering
	\includegraphics[angle=0,width=0.975\textwidth]{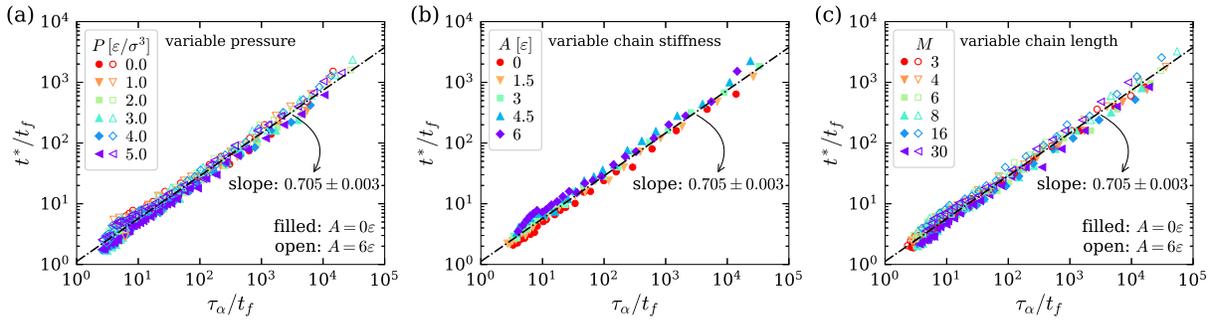}
	\caption{\label{Fig_Tau_TimeMSD}Examination of the decoupling relation between the structural relaxation time and the peak time of the non-Gaussian parameter. Panels (a--c) show $t^* / t_f$ versus $\tau_{\alpha} / t_f$ for variable $P$, $A$, and $M$, respectively. Here, $t_f$ is the fast $\beta$-relaxation time on a ps timescale. Filled and open symbols in panels (a) and (c) correspond to the results for $A = 0 \varepsilon$ and $6 \varepsilon$, respectively. Lines indicate the scaling relation, $(t^* / t_f ) \sim (\tau_{\alpha} / t_f)^{0.705 \pm 0.003}$, where the uncertainty in the exponent corresponds to the standard deviation on the fitted parameter by utilizing all the data for the fit. The analysis considers the data below the onset temperature $T_A$ of glass formation.}
\end{figure*}

We discuss the relation between the different characteristic timescales of glass formation. In previous studies of both polymeric and metallic GF liquids,~\cite{2013_JCP_138_12A541, 2015_JCP_142_164506, 2019_JCP_151_184503} it has been repeatedly shown that $t^*$ and $t_L$ correlate strongly with the lifetime of the mobile particle clusters, while $t_{\chi}$ correlates strongly with the lifetime of the immobile particles. Further, $\tau_{\alpha}$ correlates very strongly with $t_{\chi}$, and, for small-molecule and atomic liquids, $t^*$ scales in proportion to $D / k_{\mathrm{B}} T$, where $D$ is the translational diffusion coefficient. In particular, a ``decoupling'' relation between these timescales is generally observed, 
\begin{equation}
	\label{Eq_Decoupling}
	(t^* / t_f ) \sim (\tau_{\alpha} / t_f )^{1 - \zeta}
\end{equation}
where $t_f$ is the fast $\beta$-relaxation time and the exponent $\zeta$ quantifies the ``extent of decoupling''. In small-molecule liquids, $\zeta$ also quantifies the extent of the breakdown of the Stokes-Einstein relation between the diffusion coefficient and structural relaxation time.~\cite{2013_JCP_138_12A541, 2022_Mac_55_8699} It is apparent from these results that the magnitude of $\zeta$ derives from an asymmetry in the persistence time of the mobile and immobile particle clusters. Figure~\ref{Fig_Tau_TimeMSD} shows that this scaling applies to all of the materials and conditions that we simulate, where the decoupling exponent is found to be $1 - \zeta = 0.705 \pm 0.003$. This scaling relation between $t^*$ and $\tau_{\alpha}$ has repeatedly been shown before for diverse materials and appears to be a ``universal'' property of GF liquids.~\cite{2013_JCP_138_12A541, 2016_Mac_49_8355, 2017_Mac_50_2585, 2021_EPJE_44_56, 2015_JCP_142_164506, 2019_JCP_151_184503, 2022_JCP_157_064901, 2022_Mac_55_9990, 2022_JNCSX_14_100098}

It has recently been suggested that the characteristic time $t^*$ can be directly identified with the JG $\beta$-relaxation time $\tau_{\mathrm{JG}}$. We previously reviewed evidence in favor of this proposed correspondence in a previous paper,~\cite{2022_Mac_55_8699} based on relaxation data on a family of polymeric GF fluids covering a wide range of $T$ and $P$. In particular, we found that $t^*$ varies in a universal way in relation to $\tau_{\alpha}$ and that this universality derives from the insensitivity of $\zeta$ to $P$. The same ``universal'' relationship has been observed in many measurements of $\tau_{\mathrm{JG}}$ and $\tau_{\alpha}$ in a variety of GF liquids when $P$ is varied over a large range. While these observations are suggestive, they definitely do not prove $t^* = \tau_{\mathrm{JG}}$. The case for such a relationship is made somewhat more compelling by the fact that the same scaling relation as eq~\ref{Eq_Decoupling} has been observed between $\tau_{\mathrm{JG}}$ and $\tau_{\alpha}$ by Massa et al.,~\cite{2022_Polymers_14_5560} where the decoupling exponent $\zeta$ was estimated (in the notation of the present paper) to equal $1 - \zeta = 0.71 \pm 0.01$ in a coarse-grained polymer model similar to that utilized in the present work. Somewhat smaller values, $1 - \zeta = 0.69 \pm 0.01$ or $0.68 \pm 0.01$, were also reported in ref~\citenum{2022_Polymers_14_5560}, depending on the method to estimate the relaxation times. This exponent estimate is consistent within numerical uncertainty with the estimated ``decoupling'' exponent $\zeta$ obtained from the data in Figure~\ref{Fig_Tau_TimeMSD}, supporting a general quantitative correspondence between $\tau_{\mathrm{JG}}$ and $t^*$. We note that $\zeta$ for polymer fluids can be appreciably altered by additives and film confinement. Empirical evidence suggests to us that $\zeta$ is more variable in small-molecule liquids and liquid mixtures. This phenomenon clearly deserves further investigation.

We note some further evidence for the interrelation between $\tau_{\mathrm{JG}}$ and $t^*$ obtained by other researchers. Cicerone and coworkers~\cite{2014_PRL_113_117801, 2017_JCP_146_054502} have interpreted the JG $\beta$-relaxation process in terms of a particle ``hopping process'' between metabasins in the free-energy landscape, and, consistent with this hypothesis, they experimentally observed that the diffusion coefficient $D$ for a number of small-molecule GF fluids scaled inversely with $\tau_{\mathrm{JG}}$ over a wide range of $T$. Simulations of atomic GF fluids have repeatedly established that $D / k_{\mathrm{B}} T$ scales inversely to $t^*$,~\cite{2013_JCP_138_12A541, 2022_JNCSX_14_100098} so by an extension of the reasoning of Cicerone and coworkers~\cite{2014_PRL_113_117801, 2017_JCP_146_054502} we are yet again led to a correspondence between $\tau_{\mathrm{JG}}$ and $t^*$. Moreover, we mention the original study suggesting a relation between $\tau_{\mathrm{JG}}$ and $t^*$ by Leporini and coworkers,~\cite{2021_Mac_54_2053} where a direct correlation between $\tau_{\mathrm{JG}}$ and $t^*$ was found in a simulated coarse-grained polymer fluid similar to that studied in the present paper. Finally, we note that Zhang and coworkers~\cite{2021_JCP_154_084505, 2021_EPJE_44_56} found that $\tau_{\mathrm{JG}}$ correlated strongly with the lifetime $\tau_{_M}$ of the mobile particle clusters in an Al-Sm metallic GF material, which, as noted above, tracks $t^*$ in GF liquids. The accumulated circumstantial evidence supporting a correspondence between$\tau_{\mathrm{JG}}$ and $t^*$ is thus rather strong. In the future, we plan to calculate $\tau_{\mathrm{JG}}$ in our family of polymer models to more directly test the relationship between $\tau_{\mathrm{JG}}$ and $\tau_{\alpha}$ following the methods of Massa et al.~\cite{2022_Polymers_14_5560}

\subsection{\label{Sec_GOT}Characteristic Timescale of Stress Autocorrelation Function}

\begin{figure*}[htb]
	\centering
	\includegraphics[angle=0,width=0.8\textwidth]{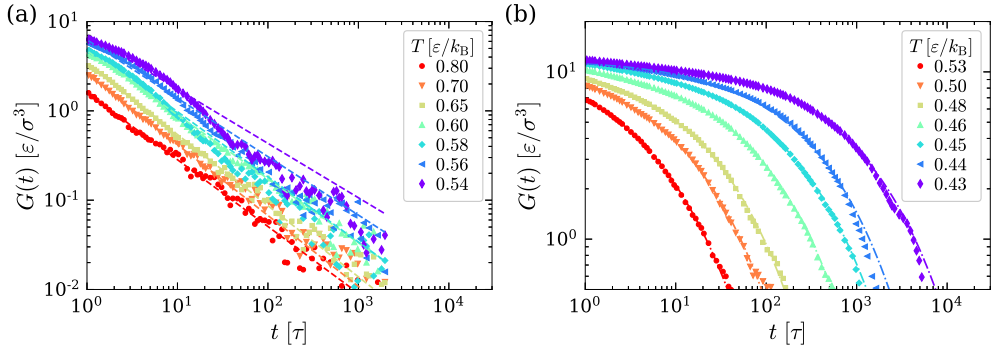}
	\caption{\label{Fig_GOT}Examination of the time dependence of the stress autocorrelation function. Panels (a) and (b) show $G(t)$ at relatively high and low $T$, respectively, for the polymer melt having $M = 20$ and $A = 0 \varepsilon$ at $P = 0.0 \varepsilon / \sigma^3$. Dashed and dashed-dotted lines in panels (a) and (b) correspond to the fits based on a power law (eq~\ref{Eq_Power}) and a stretched exponential (eq~\ref{Eq_Streched}), respectively.}
\end{figure*}

While both $G_p$ and $G_{\infty}$ are important properties associated with $G(t)$ in the short time limit, it would also be of great interest to analyze the long time behavior of $G(t)$. We thus focus on the decay of $G(t)$ at $t>1 \tau$, which provides insight into the relaxation of polymeric GF fluids. To this end, Figure~\ref{Fig_GOT} separately examines the $t$ dependence of $G(t)$ at relatively high and low $T$. Our simulation results indicate that $G(t)$ exhibits a power-law decay and a stretched exponential decay over an extended regime of $t$ in the high $T$ polymer melt regime and the low $T$ glassy segmental dynamics regime, respectively, which can be described by the following equations,
\begin{equation}
	\label{Eq_Power}
	G(t) \sim t^{-\alpha_{_G}}
\end{equation}
and
\begin{equation}
	\label{Eq_Streched}
	G(t) \sim \exp\left[ -(t/\tau_{_G})^{\beta_{_G}} \right]
\end{equation}
where $\alpha_{_G}$, $\tau_{_G}$, and $\beta_{_G}$ are fitting parameters. The fitted curves from the power-law and stretched exponential functions are shown as the dashed and dashed-dotted lines in Figures~\ref{Fig_GOT}a and~\ref{Fig_GOT}b, respectively. Since the fits are sensitive to the data quality of $G(t)$, the power-law and stretched exponential fits utilize the simulation data for $G(t)$ with the values greater than $10^{-2} \varepsilon / \sigma^3$ and $5 \times 10^{-1} \varepsilon / \sigma^3$, respectively. Moreover, the relaxation appears to be a mixed type of both power-law and stretched exponential decays at intermediate $T$. For instance, this feature can be seen at $k_{\mathrm{B}} T / \varepsilon = 0.54$ (Figure~\ref{Fig_GOT}a), where the relaxation is characterized by a power-law at short $t$ with $t < 10$, but deviations from this scaling are evident at longer $t$.

The relaxation should be Rouse-like in the high $T$ polymer melt regime so that power-law stress relaxation is expected. We may expect this type of relaxation to persist to long times when the chains are long. Relaxation in this high $T$ regime is known to be dominated by the chain dynamics, either entangled or unentangled, depending on polymer molecular mass. However, the glassy segmental dynamics begins to predominate at $T$ approaching $T_{\mathrm{g}}$ so that the nature of the relaxation process switches over to being dominated by chain segmental relaxation,~\cite{2001_JCP_114_9156} and hence, the relaxation should switch over somehow to being stretched exponential, regardless of polymer molecular mass or chemical constitution. In the glass state, the relaxation is expected to change back again to a power-law stress relaxation and ``creep compliance''~\cite{1961_RA_1_603, 1989_JPSB_27_307, 1976_PES_16_777, 1962_PM_7_2003, 1953_JMPS_1_172, 1996_PML_73_35} for all glass substances.~\cite{1991_Mac_24_3163, Book_Struik}

The emergence of power-law relaxation of $G(t)$ at low $T$ in GF liquids has important implications for understanding the nature of glass formation and the rheological properties of GF materials. Materials exhibiting this type of relaxation are in an intermediate rheological state between a Newtonian liquid and a Hookean solid in which the material stress is proportional to the strain rate and to the strain, respectively. In particular, under the ``critical conditions'', the viscosity, which is the integral of $G(t)$, becomes infinite because of the slow decay of $G(t)$. Notably, this divergence exists even though $G(t)$ decays to zero at long times and at the same time the zero-frequency shear modulus is zero. A power-law decay of $G(t)$ means that the material stress is proportional to the \textit{fractional derivative} of the strain,~\cite{Douglas_Book, 1999_JPCM_11_A329, 2022_PES_62_349} where the power of the fractional order differential operator determines the ``degree of intermediacy'' between the ordinary fluid and solid states. New transport properties are required for materials in this type of critical state at the transition from liquid-like to solid-like behavior.~\cite{2022_PES_62_349} It is not generally appreciated that this type of material is actually rather common, encompassing many gels and everyday materials, such as foods and other biological materials. From a measurement standpoint, this type of material can be problematic because the apparent viscosity or modulus can vary over a large range with the frequency.~\cite{1987_Science_235_4795} Correspondingly, we may expect relaxation times to depend appreciably on the timescale utilized to fit the relaxation data when the rheological measurements are fitted to models that are more suitable for ``ordinary'' liquids or solids.

We note that previous experimental, computational, and theoretical studies have suggested that GF materials at $T$ somewhat below $T_{\mathrm{g}}$, and, thus at $T$ below which we simulate reliably, become \textit{true elastic solids} having a finite zero-frequency shear modulus, $G_o$.~\cite{2002_PRL_89_285701, 2004_PRE_70_041501, 2010_PRL_105_015504, 2011_EPJE_34_97, 2013_JCP_138_12A533, 2017_PRL_119_147802} The emergence of a solid state with a finite zero-frequency shear modulus has important implications for the nature of relaxation in equilibrium GF materials. In particular, the stress relaxation function must acquire a universal power-law decay, mentioned above, and the other universal viscoelastic properties~\cite{1992_PRA_45_R5343} right at the \textit{dynamical critical point} where such a transition occurs. Accepting the existence of this proposed transition to an equilibrium solid state, at least in some GF materials, it then implies that the very meaning of the $\alpha$-relaxation process becomes questionable because the fluid shear viscosity is then formally infinite, as discussed above. The emergence of a solid amorphous solid state evidently has nontrivial ramifications for understanding the fundamental nature of glass formation so that further investigation is required in the future. In the present work, our discussion is confined to the \textit{equilibrium} liquid regime of incipient glass formation where the zero-frequency shear modulus is zero. It is our expectation that $G_p = G_o$ in both crystalline and amorphous solid materials in the sense just described, but this hypothesis remains to be verified.

\begin{figure*}[htb]
	\centering
	\includegraphics[angle=0,width=0.975\textwidth]{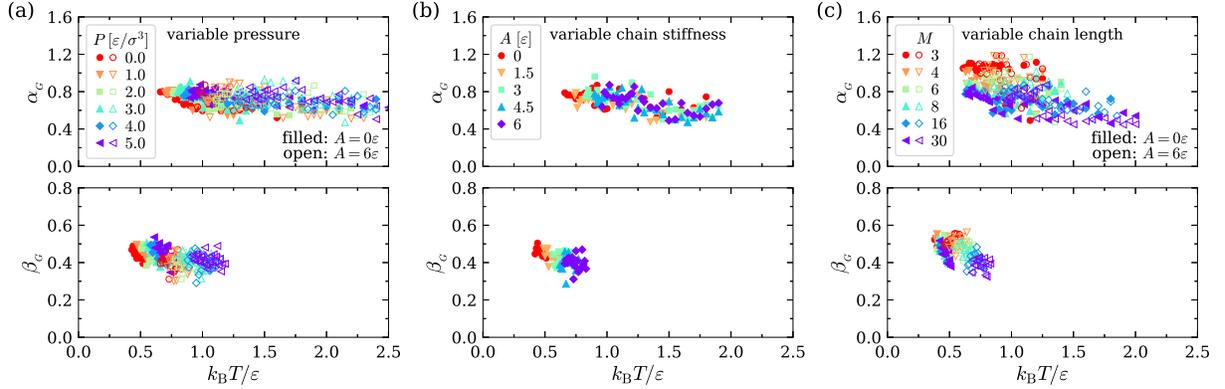}
	\caption{\label{Fig_Exponent}Temperature dependence of the exponents associated with the power-law and stretched exponential decays of $G(t)$. Panels (a--c) show $\alpha_{_G}$ and $\beta_{_G}$ versus $k_{\mathrm{B}}T / \varepsilon$ for variable $P$, $A$, and $M$, respectively. Filled and open symbols in panels (a) and (c) correspond to the results for $A = 0 \varepsilon$ and $6 \varepsilon$, respectively.}
\end{figure*}

Figure~\ref{Fig_Exponent} summarizes $T$ dependence of the exponents $\alpha_{_G}$ and $\beta_{_G}$ associated with the power-law and stretched exponential decays of $G(t)$, respectively. Here, the analysis is made for a range of polymer melts having variable pressure $P$, chain stiffness $A$, and chain length $M$. Specifically, the chain length is fixed at $M = 20$ for variable $P$ and $A$ and the pressure is fixed at $P = 0.0 \varepsilon / \sigma^3$ for variable $M$. For the range of $T$ considered, both $\alpha_{_G}$ and $\beta_{_G}$ seem to be nearly independent of $T$. In particular, $\alpha_{_G}$ falls in a narrow range between $0.4$ and $1.2$, as thermodynamic ($T$ and $P$) and molecular ($A$ and $M$) variables are altered. We also find that the ``stretching exponent'' $\beta_{_G}$ is weakly dependent on $T$ and falls in the range of $0.3$ to $0.6$. We note that $\beta_{_G}$ has often been observed to be relatively constant under moderate cooling in GF liquids, but $\beta_{_G}$ tends to decrease upon lowering $T$ towards $T_{\mathrm{g}}$ in the low temperature regime of glass formation below the crossover temperature $T_c$.~\cite{2009_JCP_130_124902, 2017_JTAC_127_1975} Unfortunately, this low $T$ regime of glass formation below $T_c$ is inaccessible to the current simulations. Nevertheless, in rough consistency with measurement, $\beta_{_G}$ is observed to be roughly constant in the $T$ range studied by our simulations.

The $T$ dependence of the stretching exponent $\beta_{\mathrm{ISF}}$ describing the long time relaxation of $F_s(q^*, t)$ over a wide range of $T$ has been carefully described in a previous work.~\cite{2020_SciAdv_6_eaaz0777} The determination of this exponent is made easier for this property by the lack of the large fluctuations in $G(t)$ in comparison to $F_s(q^*, t)$. These fluctuations are apparent in the fast dynamics regime of $G(t)$ in Figure~\ref{Fig_TCF}. $F_s(q^*, t)$ has been subjected to an extensive analysis in ref~\citenum{2020_SciAdv_6_eaaz0777} for a coarse-grained polymer model similar to that utilized in the present paper, so we do not repeat this analysis here for reasons of space. We briefly point out that this previous analysis indicates that there are \textit{two} $\beta$ exponents and relaxation times describing the multi-step decay of $F_s(q^*, t)$ where the first exponent $\beta_f$ and relaxation time $\tau_f$ are nearly invariant to $T$, while the exponent $\beta_{\mathrm{ISF}}$ governing the longer $\alpha$-relaxation time monotonically decreases in a smooth fashion as $T$ is reduced from a high $T$ regime where the $\alpha$- and $\beta$-relaxation processes merge approximately at the onset temperature $T_A$ to relatively low values at the lowest $T$ that could be simulated.~\cite{2020_SciAdv_6_eaaz0777} In future work, we plan to check whether or not there is any direct relationship between the exponent $\beta_{\mathrm{ISF}}$ and the extent $L$ of collective motion in our GF polymer fluids, as some have heuristically suggested. At this point, it seems clear that the constancy of the decoupling exponent $\zeta$ in eq~\ref{Eq_Decoupling} would preclude any relationship to $\beta_{\mathrm{ISF}}$, which clearly depends appreciably on $T$. These observations lead us to suggest that $\zeta$ might be instead related to the intermittent ``jump-like'' particle displacement events that characterize the molecular dynamics of the JG relaxation process.~\cite{2021_EPJE_44_56, 2021_JCP_154_084505} Zhang et al.~\cite{2021_JCP_154_084505} found in a study of JG relaxation that these jumps occurring on a ``fractal time'' set a dimension of $0.4$, based on a model Al-Sm metallic glass material, a phenomenon that naturally explains the Cole-Cole-type relaxation commonly observed for the JG relaxation process.~\cite{1999_JPCM_11_A329, 2021_JCP_154_084505} In polymer materials, we may expect the intermittent jump motions to predominantly involve local rotational or chain torsional motions~\cite{2005_PRE_71_050801} rather than molecular translational jumps that predominate the particle jump motion in a metallic glass material. If this explanation of the decoupling exponent is confirmed, it would also have implications for the interpretation of $\beta_{_G}$. We hope to test this possibility in the near future.

\begin{figure*}[htb]
	\centering
	\includegraphics[angle=0,width=0.975\textwidth]{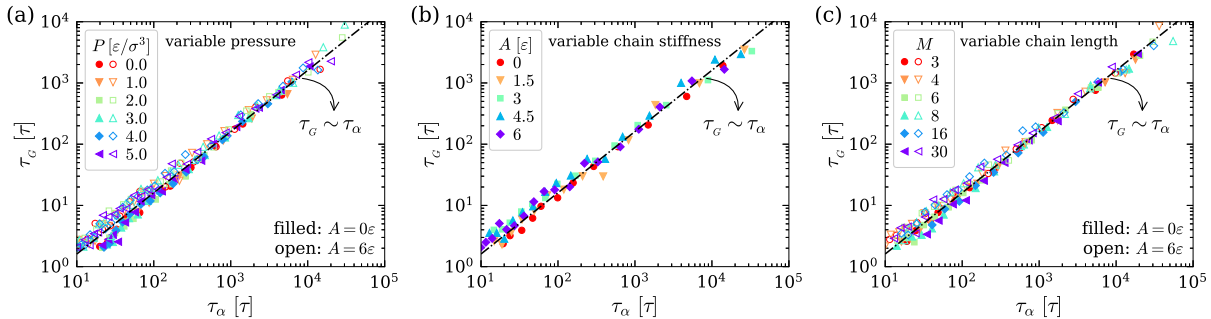}
	\caption{\label{Fig_Tau_TauG}Correlation between the relaxation times determined from the self-intermediate scattering function and stress autocorrelation function. Panels (a--c) show $\tau_{_G}$ versus $\tau_{\alpha}$ for variable $P$, $A$, and $M$, respectively. Filled and open symbols in panels (a) and (c) correspond to the results for $A = 0 \varepsilon$ and $6 \varepsilon$, respectively. Lines indicate the direct proportionality, $\tau_{_G} \sim \tau_{\alpha}$. The analysis considers the data in the relatively low $T$ regime.}
\end{figure*}

Regarding the $T$ dependence of $\tau_{_G}$, our analysis reveals that this characteristic timescale closely tracks $\tau_{\alpha}$. In particular, Figure~\ref{Fig_Tau_TauG} indicates the presence of a near direct proportionality between the two timescales in the low $T$ regime where the glassy segmental dynamics dominates the stress relaxation. Remarkably, the scaling relation, $\tau_{_G} \sim \tau_{\alpha}$, holds in this regime for all the polymer models that we consider. It is this scaling relation that justifies referring to $\tau_{\alpha}$ as the ``structural relaxation time'' in fluids approaching their glass transition temperatures where the chain connectivity contribution to the relaxation process becomes almost irrelevant.~\cite{2001_JCP_114_9156} We suggest that the near constancy of the segmental relaxation time of unentangled polymer liquids and many molecular liquids, i.e., $\tau_{\alpha}(T_{\mathrm{g}}) \sim 100$ s, is responsible for the remarkable relative constancy of the shear viscosity of unentangled polymer liquids near $T_{\mathrm{g}}$, i.e., $\eta(T_{\mathrm{g}}) \sim 10^{12}\ \mathrm{Pa} \cdot \mathrm{s}$.~\cite{1968_JPSA_6_249} A new ``terminal relaxation process'' arises in entangled polymer melts at times much longer than the segmental relaxation time, and correspondingly, $\eta$ at $T_{\mathrm{g}}$ becomes mass dependent in entangled polymer fluids. Again, we emphasize that the present simulation study is limited to unentangled polymer melts.

\subsection{\label{Sec_Elastic}Test of Elastic and String Models of Glass Formation}

The elastic models of glass formation relate the increasing solidity of GF materials to the dramatic slowing down of the structural relaxation and the viscous flow. To understand the physical significance of $G_p$, we begin by examining to what extent the structural relaxation can be described by the glassy plateau shear modulus. The key prediction of the elastic models is that the structural relaxation is activated with an energy barrier $\Delta E$, due to the cage breaking, being proportional to an elastic modulus which is explicitly identified with the ``instantaneous shear modulus'' $G_{\mathrm{inst}}$,~\cite{2006_RMP_78_953, 2012_JCP_136_224108, 2015_JNCS_407_14}
\begin{equation}
	\label{Eq_Elastic}
	\tau_{\alpha} = \tau_o \exp(\Delta E / k_{\mathrm{B}} T),\ \Delta E = G_{\mathrm{inst}} V_c^*
\end{equation}
where $\tau_o$ is a prefactor and $V_c^*$ is an empirical ``characteristic volume'' that is normally assumed to be on the order of the molecular or polymer segment volume and to be independent of $T$. Puosi and Leporini~\cite{2012_JCP_136_041104} first emphasized that $G_{\mathrm{inst}}$ should not be identified with the infinite frequency shear modulus, $G_{\infty}$, whose thermodynamic definition is well known, but rather should be identified with a glassy ``plateau modulus'' $G_p$, corresponding to $G(t)$ after an initial fast $\beta$ stress relaxation process, as in the case of the self-intermediate scattering function (see Figure~\ref{Fig_TCF}a). Since the fast $\beta$-relaxation time $t_f$ is generally on the order of $1$ ps for both molecular and atomic GF liquids,~\cite{1997_JPCM_9_10079, 1997_PTPS_126_159, 2018_JCP_148_104508} we correspondingly define $G_p$ at $t = 1 \tau$, as mentioned in Section~\ref{Sec_Definition}. Notably, $G_p$ cannot be calculated from any known thermodynamic relation. As we shall illustrate below, $G_p$ has a \textit{qualitatively} different $T$ dependence than $G_{\infty}$. This qualitative difference has been noted before,~\cite{2020_SciAdv_6_eaaz0777, 2012_JCP_136_041104} and Dyre and coworkers~\cite{2012_JCP_136_224108} accordingly modified the ``shoving model'' by replacing $G_{\infty}$ by $G_p$.

Puosi and Leporini~\cite{2012_JCP_136_041104} further proposed an interesting specific relation between $\tau_{\alpha}$ and $G_p$ that has been verified for a coarse-grained model of polymeric and small-molecule GF liquids,
\begin{equation}
	\label{Eq_Quadratic}
	\ln (\tau_{\alpha}) = c_0 + c_1 (G_p / k_{\mathrm{B}} T) + c_2 (G_p / k_{\mathrm{B}} T)^2
\end{equation}
where $c_0$, $c_1$, and $c_2$ are adjustable constants. We consider this interesting relation below for a different family of coarse-grained polymer models, where the polymer stiffness and applied pressure are varied over a large range to further test the generality of this relation.

\begin{figure*}[htb]
	\centering
	\includegraphics[angle=0,width=0.975\textwidth]{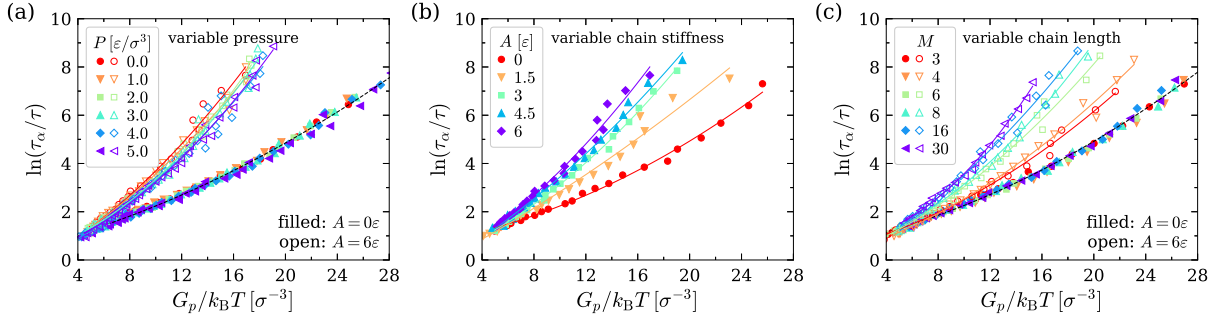}
	\caption{\label{Fig_GpT_Tau}Correlation between the glassy plateau shear modulus and the structural relaxation time. Panels (a--c) show $\ln (\tau_{\alpha} / \tau)$ versus $G_p / k_{\mathrm{B}}T$ for variable $P$, $A$, and $M$, respectively. Filled and open symbols in panels (a) and (c) correspond to the results for $A = 0 \varepsilon$ and $6 \varepsilon$, respectively. Lines are a guide to the eye. The analysis considers the data below $T_A$.}
\end{figure*}

Figure~\ref{Fig_GpT_Tau} examines the correlation between $\ln (\tau_{\alpha})$ versus $G_p / k_{\mathrm{B}}T$ in our models of polymer melts having variable pressure, chain rigidity, and chain length. First, consistent with the results of Puosi and Leporini,~\cite{2012_JCP_136_041104} our results indicate the occurrence of a near universal relation between $\tau_{\alpha}$ and $G_p / k_{\mathrm{B}}T$ for polymer melts having variable chain length in the absence of bending constraints, as shown in Figure~\ref{Fig_GpT_Tau}c. While the quadratic functional form appears to apply to all the polymer materials that we consider, the associated constants (i.e., $c_0$, $c_1$, and $c_2$) can depend appreciably on chain rigidity and other molecular parameters. We find similar results if we replace $G_p / k_{\mathrm{B}} T$ by $\langle u^2 \rangle_o / \langle u^2 \rangle$ where $\langle u^2 \rangle_o$ is the value of $\langle u^2 \rangle$ at a reference temperature, as one would naturally expect from the near linear relation between these quantities that we discuss below. The original formulation of Leporini and coworkers~\cite{2008_NaturePhys_4_42} involved exactly this type of relation between $\tau_{\alpha}$ and stiffness defined in terms of $\langle u^2 \rangle_o / \langle u^2 \rangle$. We may then conclude that the activation energy $\Delta G$ for structural relaxation increases progressively with increasing molecular rigidity, a basic premise of both the ``shoving model''~\cite{2006_RMP_78_953, 2012_JCP_136_224108, 2015_JNCS_407_14} and the family of models proposed by Leporini and coworkers,~\cite{2008_NaturePhys_4_42} which are built on this conceptual model of the origin of the growth of the structural relaxation time in cooled liquids. We do not discuss here the localization model of glass formation,~\cite{2012_SoftMatter_8_11455, 2015_PNAS_112_2966, 2016_JSM_054048} which likewise predicts a relation between $\Delta G$ and $\langle u^2 \rangle$, because the fitting to the model does not alter the conclusions here based on the glassy dynamics relaxation model of Leporini and coworkers.~\cite{2008_NaturePhys_4_42} This alternative model describes the evident nonlinear relation between $\Delta G$ and $1 / \langle u^2 \rangle$, first observed in the simulations of Starr et al.,~\cite{2002_PRL_89_125501} by a power-law rather than a Taylor series expansion truncated at the second order, as in the original model of Leporini and coworkers relating $\Delta G$ to $1 / \langle u^2 \rangle$.~\cite{2008_NaturePhys_4_42} 

\begin{figure*}[htb]
	\centering
	\includegraphics[angle=0,width=0.975\textwidth]{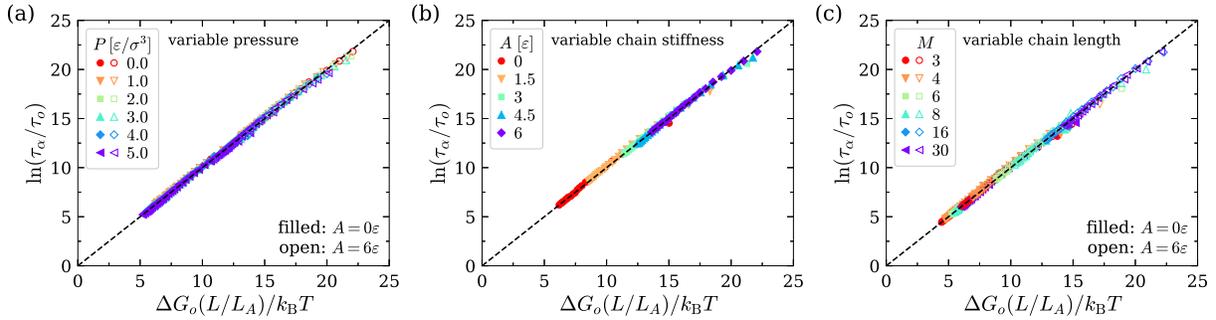}
	\caption{\label{Fig_String}String model description of glass formation. Panels (a--c) show $ \Delta G_o(L/L_A)/k_{\mathrm{B}}T$ versus $\ln (\tau_{\alpha}/\tau_o)$ for variable $P$, $A$, and $M$, respectively. Filled and open symbols in panels (a) and (c) correspond to the results for $A = 0 \varepsilon$ and $6 \varepsilon$, respectively. Dashed lines indicate $\ln (\tau_{\alpha}/\tau_o) = \Delta G_o(L/L_A)/k_{\mathrm{B}}T$, where $\tau_o$, $L_A$, and $\Delta G_o$ are explained in the text.}
\end{figure*}

As a backdrop to frame our investigation below, we also present a test of the string model of glass formation~\cite{2014_JCP_140_204509, 2014_JCP_141_141102} to quantify the extent of cooperative motion in our model GF liquids. This model is built on the underlying framework of TST~\cite{1941_CR_28_301, Book_Eyring} and is broadly consistent with the entropy theory of glass formation.~\cite{1965_JCP_43_139, 2021_Mac_54_3001} To describe the dynamics at temperatures below the ``onset temperature'' $T_A$ where the $\alpha$-relaxation process is non-Arrhenius, the string model of glass formation assumes that the activation free energy $\Delta G$ for structural relaxation is proportional to the average string length $L$ normalized by its value at the onset temperature $T_A$, leading to a $T$-dependent activation free energy as $\Delta G(T) = \Delta G_o (L/L_A)$, where $L_A$ is the value of $L$ at $T_A$. This, in turn, yields the following expression for $\tau_{\alpha}$,
\begin{eqnarray}
	\label{Eq_String}
	\tau_{\alpha} = \tau_{o} \exp\left( \frac{\Delta G_o}{k_{\mathrm{B}}T} \frac{L}{L_A} \right)
\end{eqnarray}
The parameter $\tau_{o}$ can be eliminated from a knowledge of $\tau_{\alpha}$ at $T_A$,~\cite{2015_PNAS_112_2966} and $\Delta H_o$ may be determined from the Arrhenius equation in the high $T$ regime where standard TST is assumed to be applicable as a descriptive framework for liquid dynamics, resulting in the following equation with $\Delta S_o$ being the only fitting parameter,
\begin{eqnarray}
	\tau_{\alpha} = \tau_{\alpha} (T_A)\exp\left(\frac{\Delta H_o - T\Delta S_o}{k_{\mathrm{B}}T}\frac{L}{L_A} - \frac{\Delta H_o - T_A\Delta S_o}{k_{\mathrm{B}}T_A}\right)
\end{eqnarray}
We use the above equation as a method to test the validity of the string model of glass formation and estimate $\Delta S_o$ if our simulation results conform to the string model of glass formation. Figure~\ref{Fig_String} shows that all our simulation data can be satisfactorily described by the above equation in our coarse-grained polymer models having variable $P$, $A$, and $M$.

\begin{figure*}[htb!]
	\centering
	\includegraphics[angle=0,width=0.975\textwidth]{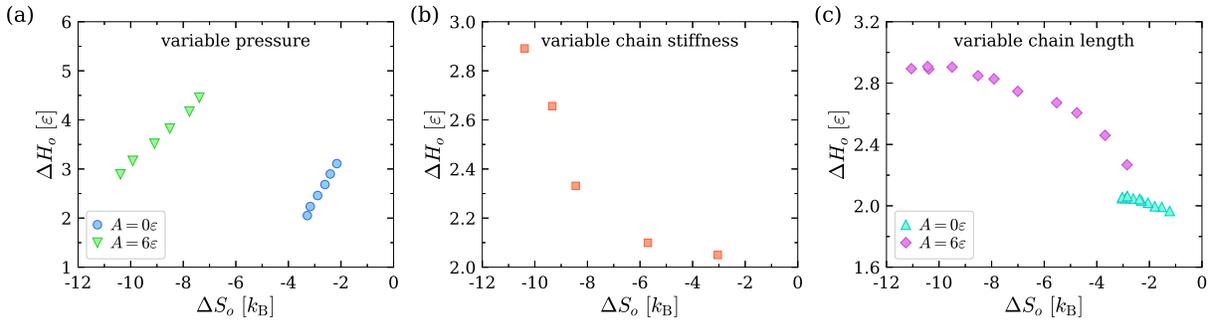}
	\caption{\label{Fig_StringHS}Correlation between the enthalpy $\Delta H_o$ and entropy $\Delta S_o$ of high temperature activation for polymer melts having various variables. Panels (a--c) correspond to the results for variable $P$, $A$, and $M$, respectively.}
\end{figure*}

For completeness, Figure~\ref{Fig_StringHS} shows the correlation between $\Delta H_o$ and $\Delta S_o$ for polymer melts having variable applied pressure, chain stiffness, and chain length. In previous simulations of glass formation of various polymeric systems,~\cite{2014_NatCommun_5_4163, 2015_PNAS_112_2966, 2015_JCP_142_234907} $\Delta H_o$ and $\Delta S_o$ have been found to vary in a proportional manner, so that these two parameters exhibit the entropy-enthalpy compensation effect. Reference~\citenum{2015_JCP_143_144905} has provided a detailed discussion on this phenomenon in polymer liquids and Gelin et al.~\cite{2020_NatCommun_11_3977} have recently discussed this important phenomenon from a normal mode perspective and the classical transition state theory of condensed state dynamics in the case of metallic GF liquids. Our previous simulation works have also demonstrated the occurrence of strong positive correlations between $\Delta H_o$ and $\Delta S_o$ in polymer melts having variable cohesive energy~\cite{2020_Mac_53_9678} and pressure,~\cite{2020_Mac_53_6828} but a negative correlation between these energetic parameters has also been observed in polymer melts with varying chain stiffness.~\cite{2020_Mac_53_4796} Figure~\ref{Fig_StringHS} indicates the presence of a strong correlation between $\Delta H_o$ and $\Delta S_o$ when varying individual thermodynamic or molecular parameters, despite the absence of a universal behavior.

We conclude from the above analyses that the activation free energy $\Delta G(T)$ increases with the extent of collective motion, as quantified by $L$, and the material stiffness, as quantified by $G_p$ and $\langle u^2 \rangle_o / \langle u^2 \rangle$, as temperature is lowered, pressure is increased, and chain stiffness is increased. However, the scaling relationship between these quantities is clearly not linear. We also observe that $\Delta G(T)/ \Delta G_o$ appears to equal $L / L_A$ to a high degree of approximation~\cite{2014_JCP_140_204509, 2020_Mac_53_6828} under all conditions simulated so that the relation between the scale of collective motion and $\Delta G(T)$ seems to be mathematically simpler than that between $\Delta G(T)$ and $G_p$. On the other hand, measurements of material stiffness are currently much easier than measurements of $L$ and $\langle u^2 \rangle$, which makes formulations of relaxation in terms of material stiffness highly advantageous from a practical standpoint.

\subsection{\label{Sec_Correlation}Correlation between Emergent Stiffness $G_p$ and Collective Motion}

\begin{figure*}[htb]
	\centering
	\includegraphics[angle=0,width=0.975\textwidth]{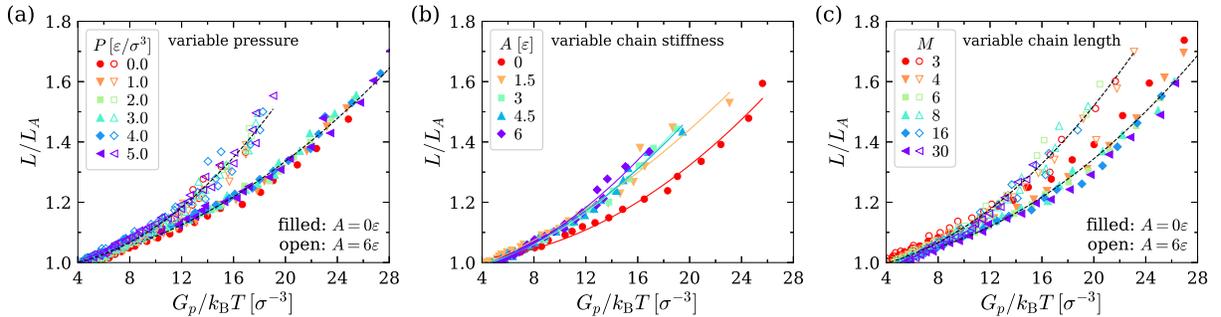}
	\caption{\label{Fig_GpT_CRR}Correlation between the glassy plateau shear modulus and the extent of collective motion. Panels (a--c) show $L / L_A$ versus $G_p / k_{\mathrm{B}}T$ for variable $P$, $A$, and $M$, respectively. Filled and open symbols in panels (a) and (c) correspond to the results for $A = 0 \varepsilon$ and $6 \varepsilon$, respectively. Lines are a guide to the eye. The analysis considers the data below $T_A$.}
\end{figure*}

Douglas and coworkers~\cite{2015_PNAS_112_2966} have examined the models of glass formation emphasizing material elasticity, free volume, and the extent of collective motion in a class of model polymeric GF liquids, such as polymer nanocomposites and polymer thin films, where the $T$ dependence of the dynamics (i.e., fragility) can be ``tuned'' over a large range by varying the nanoparticle concentration or film thickness. This analysis indicates that the dynamical free volume $\langle u^2 \rangle^{3/2}$, emergent elasticity, and collective motion models of glass formation lead to largely equivalent mathematical descriptions of the $T$ dependence of $\tau_{\alpha}$ when these quantities are determined in terms of well-defined measures of these physical characteristics of GF liquids, suggesting deep relations between these superficially disparate quantities. In this fashion, this work uncovered some unity in our understanding of GF materials from perspectives considered formerly as distinct. This has also been emphasized in our recent works,~\cite{2022_Mac_55_8699, 2023_CJPS_Xu} and the present work continues to amplify on this working hypothesis. Following this logical thread further, we examine the correlation between $G_p$ and $L / L_A$. As shown in Figure~\ref{Fig_GpT_CRR}, we find that $G_p / k_{\mathrm{B}} T$ correlates strongly with $L / L_A$ in the family of GF polymer fluids considered.

\subsection{\label{Sec_OtherCorrelation}Correlation between Glassy Plateau Modulus, Debye-Waller Parameter, and Non-Gaussian Parameter}

\begin{figure*}[htb]
	\centering
	\includegraphics[angle=0,width=0.975\textwidth]{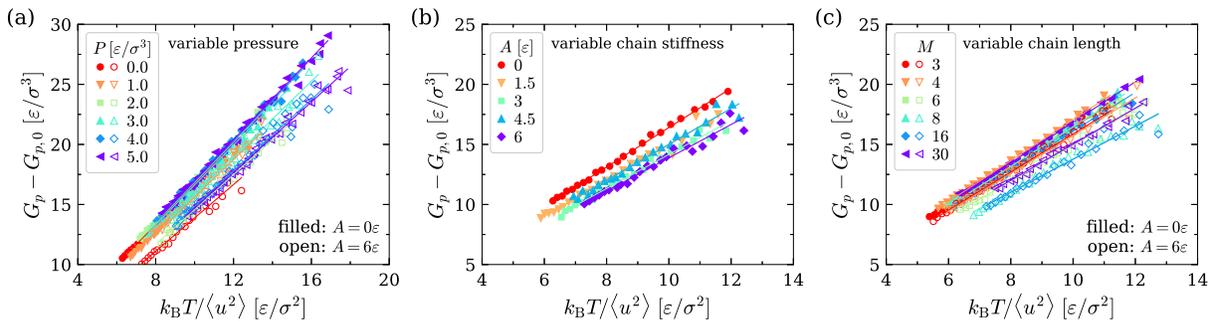}
	\caption{\label{Fig_Gp_DWF}Correlation between the glassy plateau shear modulus and the Debye-Waller parameter. Panels (a--c) show $G_p - G_{p,0}$ versus $k_{\mathrm{B}}T / \langle u^2 \rangle$ for variable $P$, $A$, and $M$, respectively. Filled and open symbols in panels (a) and (c) correspond to the results for $A = 0 \varepsilon$ and $6 \varepsilon$, respectively. Lines are a guide to the eye. The analysis considers the data below $T_A$.}
\end{figure*}

\begin{figure*}[htb]
	\centering
	\includegraphics[angle=0,width=0.975\textwidth]{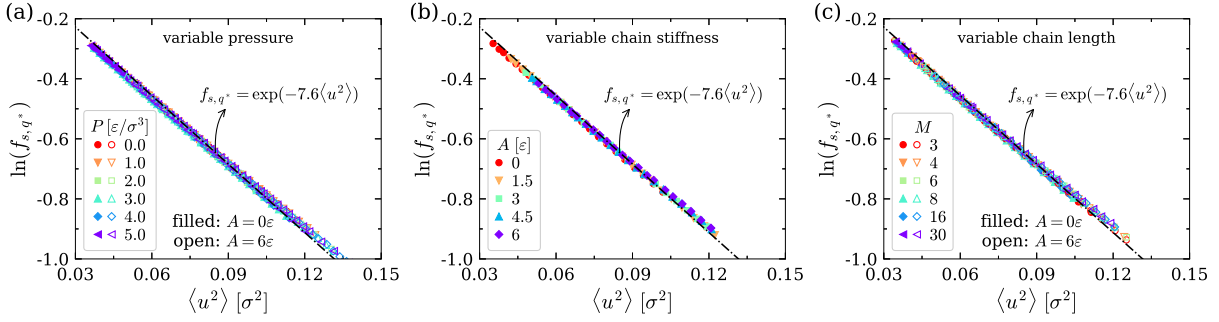}
	\caption{\label{Fig_NEP_DWF}Relation between the non-ergodicity parameter and the Debye-Waller parameter. Panels (a--c) show $\ln(f_{s, q^*})$ versus $\langle u^2 \rangle$ for variable $P$, $A$, and $M$, respectively. Filled and open symbols in panels (a) and (c) correspond to the results for $A = 0 \varepsilon$ and $6 \varepsilon$, respectively. Lines indicate the scaling relation, $f_{s, q^*} = \exp (-7.6 \langle u^2 \rangle)$. The analysis considers the data below $T_A$.}
\end{figure*}

The ``shoving'' model relating $\tau_{\alpha}$ to the material stiffness $G_p$ is based on an additional finding, which provides an important component of this model, and has proven to be independently important in other contexts. In particular, Leporini and coworkers~\cite{2015_EPJE_38_87, 2012_JCP_136_041104} proposed an approximate linear relation between $G_p$ and $k_{\mathrm{B}} T / \langle u^2 \rangle$,
\begin{equation}
	\label{Eq_GpDWF}
	G_p \approx G_{p,0} + G_{p,1} (k_{\mathrm{B}} T / \langle u^2 \rangle)
\end{equation}
where $G_{p,0}$ and $G_{p,1}$ are material-dependent constants. These authors, and others since their work, have confirmed this relation for many different types of materials.~\cite{2015_EPJE_38_87, 2021_JCP_155_204504, 2022_JCP_157_064901, 2022_Mac_55_9990} This relation is remarkable because it relates a local stiffness measure to the macroscopic stiffness of the material. We find in Figure~\ref{Fig_Gp_DWF} that eq~\ref{Eq_GpDWF} also holds very well in our polymer models having variable pressure, chain rigidity, chain length, and temperature, although the constants $G_{p,0}$ and $G_{p,1}$ are material and pressure dependent. We attribute this variability to anharmoinic interactions as $G_{p,0} = 0$ and $G_{p,1}$ takes a fixed value in the idealized Debye equation of state model of condensed materials, where the molecules exhibit idealized harmonic intermolecular interactions. Evidently, $G_{p,0}$ and $G_{p,1}$ provide valuable information about these anharmonic interactions, which remains to be understood and utilized for other purposes. Saw and Harrowell~\cite{2016_PRL_116_137801} have suggested that a universal relation exists between $G_p / G_{\infty}$ and $\langle u^2 \rangle$, regardless of whether the system is a GF liquid or a crystal. In the present work, we consider a special limit of their result corresponding to a ps timescale at which $G_p$ and $\langle u^2 \rangle$ are both defined.

Correspondingly, ref~\citenum{2018_JCP_148_104508} has discussed a general relation between another basic stiffness measure $f_{s, q^*}$ and $\langle u^2 \rangle$, which appears to also hold regardless of whether the material is crystalline or amorphous. Consistent with the result of ref~\citenum{2018_JCP_148_104508}, Figure~\ref{Fig_NEP_DWF} also indicates that $f_{s, q^*}$ for all the systems that we simulate is well described by a Gaussian function, $f_{s, q^*} = \exp(- C \langle u^2 \rangle)$, where the constant $C = 7.6$ is not exactly equal to $(q^*)^2 / 6 = 49 / 6 \approx 8.17$ expected for an ideal Gaussian atomic displacement process. The deviation from a Gaussian intermediate scattering function on a ps timescale has been emphasized by Cicerone and coworkers.~\cite{2014_PRL_113_117801} In particular, their work provides experimental evidence for ``tightly caged'' molecules having relatively small $\langle u^2 \rangle$ values and ``loosely caged'' molecules whose $\langle u^2 \rangle$ values are much larger.

\begin{figure*}[htb]
	\centering
	\includegraphics[angle=0,width=0.975\textwidth]{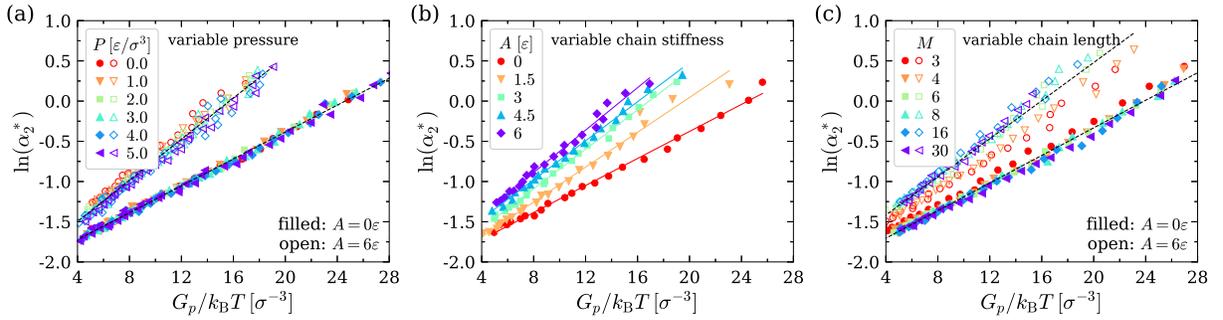}
	\caption{\label{Fig_GpT_PeakMSD}Correlation between the glassy plateau shear modulus and the non-Gaussian parameter. Panels (a--c) show $\ln (\alpha_2^*)$ versus $G_p / k_{\mathrm{B}}T$ for variable $P$, $A$, and $M$, respectively. Filled and open symbols in panels (a) and (c) correspond to the results for $A = 0 \varepsilon$ and $6 \varepsilon$, respectively. Lines are a guide to the eye. The analysis considers the data below $T_A$.}
\end{figure*}

We also check the generality of a relation between the peak height $\alpha_2^*$ of the non-Gaussian parameter and $G_p$ identified in a previous paper.~\cite{2022_Mac_55_8699} Figure~\ref{Fig_GpT_PeakMSD} examines the correlation between $G_p / k_{\mathrm{B}}T$ and $\alpha_2^*$. A near linear relation appears to hold between $G_p / k_{\mathrm{B}}T$ and $\ln (\alpha_2^*)$ in all cases. This striking relation between a generally accepted measure of ``dynamic heterogeneity'' and the effective stiffness of the material serves as a reminder that many aspects of GF liquids remain poorly understood. In particular, it is not exactly clear how this quantity actually relates physically to the various types of observed specific forms of dynamic heterogeneity observed generally in GF materials, despite the widespread consideration of $\alpha_2 (t)$ in simulations and measurements on GF liquids. At present, we simply do not have any theoretical framework for understanding why the maximum in $\alpha_2 (t)$ should be related to the material stiffness or any specific type of dynamic heterogeneity. We suspect that a better understanding of this relation could lead to an enhanced understanding of glass formation broadly, but we leave this task for future work.

Finally, we note that the amplitude of the fast relaxation process, as quantified by $1 - f_{s, q^*}$, is also of great interest from an experimental standpoint. Zhang et al.~\cite{2021_JCP_154_084505} have found that this quantity can be described approximately by the relation $1 - f_{s, q^*} \sim \langle u^2 \rangle$ because of the relation of $f_{s, q^*}$ to $\langle u^2 \rangle$ discussed in ref~\citenum{2018_JCP_148_104508}. Kojima et al.~\cite{1997_JPCM_9_10079} have discussed the measurement of the intensity of the fast relaxation, where the intensity of the fast relaxation was found to scale linearly with $T$ in the low $T$ regime, consistent with the $T$ variation of $\langle u^2 \rangle$. Betancourt et al.~\cite{2018_JCP_148_104508} have shown that the magnitude of $\langle u^2 \rangle$ in a model GF liquids similar to those studied in the present paper, thus the magnitude of the quasi-elastic neutron scattering intensity, is also dominated by string-like collective motion occurring on a ps timescale. Although this collective motion in the fast dynamics regime bears some geometrical resemblance to the string-like collective motion discussed above in connection to understanding the $T$ dependence of $\Delta G$, this type of collective motion has the rather distinct property of growing upon \textit{heating} rather than cooling. This is natural because the fast dynamics on a ps timescale is dominated by the inertial dynamics of the fluid rather than diffusion processes which tend to operate on much longer timescales, especially at low $T$. String-like collective motion on a ps timescale, with an ``increasing probability of the number of particles that participate in this process at a given time'' upon heating, has been reported in low-$q$ inelastic coherent neutron scattering measurements on a model fragile GF material.~\cite{2000_PRL_84_3630} String-like collective motion, where the motion is reversible rather than irreversible, corresponding to stable anharmonic collective modes rather than irreversible particle displacements, have been shown to give rise to a boson peak in the density of states.~\cite{2021_JCP_154_084505, 2022_NatPhys_18_669} Glass formation evidently involves a hierarchy of dynamical heterogeneity processes of significance for a comprehensive understanding of the dynamics of GF liquids.

\subsection{\label{Sec_Scaling}Consistency of Properties with Thermodynamic Scaling}

As discussed in the introduction, numerous experimental and computational studies~\cite{2005_RPP_68_1405, 2010_Mac_43_7875, Book_Roland, Book_Paluch} have established that most liquids seem to exhibit another remarkable, yet poorly understood, property termed ``thermodynamic scaling'' in which the structural relaxation time $\tau_{\alpha}$, and many other dynamic properties, can be expressed in terms of a ``universal'' reduced variable, $TV^{\gamma_t}$, where $\gamma_t$ is a scaling exponent describing how $T$ and $V$ are linked to each other when either quantity is varied.

\begin{figure*}[htb]
	\centering
	\includegraphics[angle=0,width=0.975\textwidth]{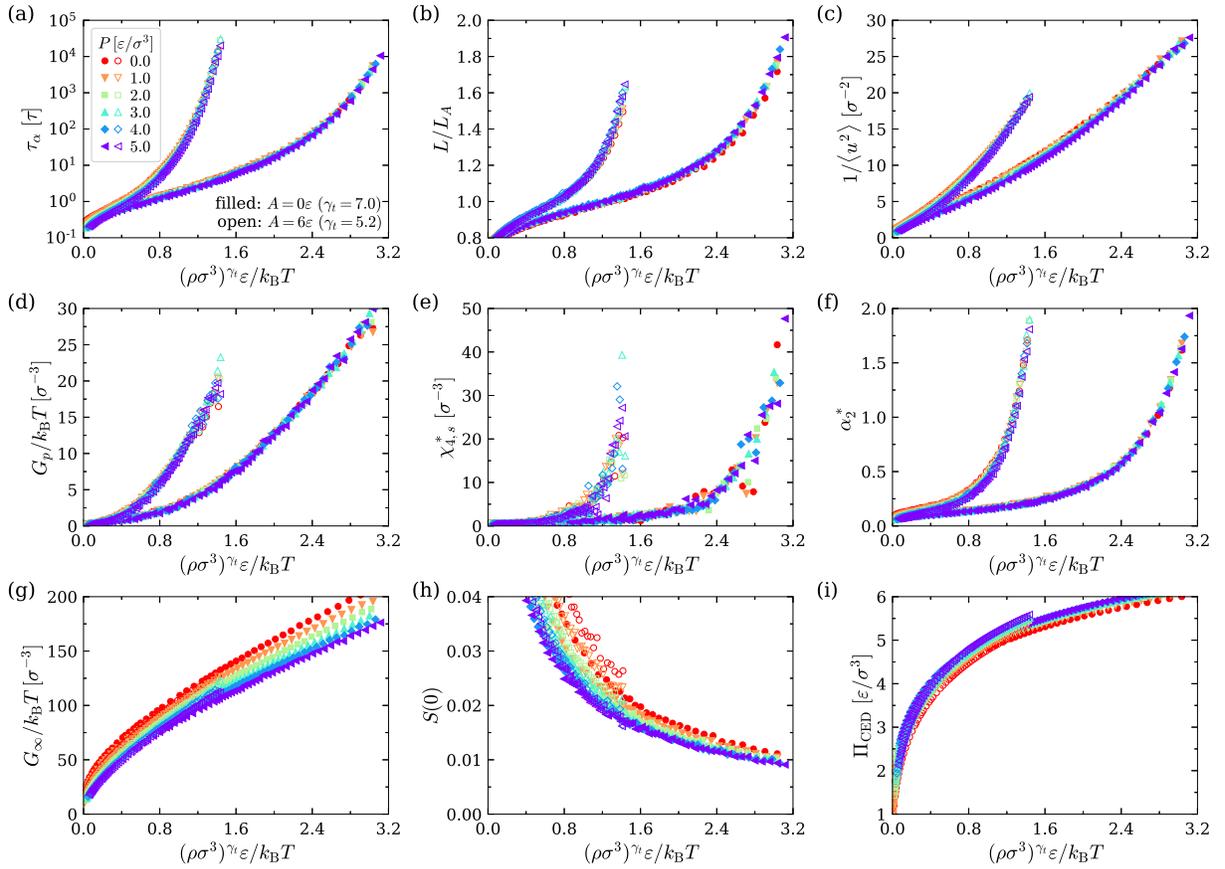}
	\caption{\label{Fig_Scaling}Test of thermodynamic scaling of various properties. Panels (a--i) correspond to the results for $\tau_{\alpha}$, $L / L_A$, $1 / \langle u^2 \rangle$, $G_p$, $\chi_{4,s}^*$, $\alpha_2^*$, $G_{\infty} / k_{\mathrm{B}} T$, $S(0)$, and $\Pi_{\mathrm{CED}}$, respectively. Filled and open symbols correspond to the results for $A = 0 \varepsilon$ and $6 \varepsilon$, respectively, for which the scaling exponents $\gamma_t$ are determined to be $7.0$ and $5.2$. $\chi_{4,s}^*$ exhibits some non-universality, but the deviations are apparently smaller than for $S(0)$. This breakdown of thermodynamic scaling of $\chi_{4,s}^*$ is discussed in a previous paper.~\cite{2021_Mac_54_3247} This non-universality has been seen experimentally.}
\end{figure*}

We now speculate about the origin of ``thermodynamic scaling'' of the glassy plateau shear modulus and the absence of this scaling for other properties. Leporini and coworkers~\cite{2012_JCP_136_041104, 2016_JCP_145_234904, 2017_JPCM_29_135101} have emphasized an aspect of the thermodynamic scaling of $G_p$ which might be crucially important for its occurrence. They observed that thermodynamic scaling arises in the glassy plateau $G_p$ divided by $k_{\mathrm{B}} T$ and in $\langle u^2 \rangle$,~\cite{2016_JCP_145_234904, 2017_JPCM_29_135101} which they rationalized based on earlier arguments by Tobolsky~\cite{Book_Tobolsky} that $G_p$ should be predominated by \textit{intermolecular} interactions so that this property is related to the cohesive energy density $\Pi_{\mathrm{CED}}$ of the liquid. Correspondingly, atomic motions not involving bond displacements clearly dominate the magnitude of $\langle u^2 \rangle$, and Leporini and coworkers found $\langle u^2 \rangle$ to be directly related to $G_p$.~\cite{2016_JCP_145_234904, 2017_JPCM_29_135101} We have confirmed these results in our own simulations, as discussed in Section~\ref{Sec_Correlation}.

In a previous work,~\cite{2021_Mac_54_3247} we have shown that thermodynamic scaling of the relaxation time can be formally derived by combining the Murnaghan equation of state~\cite{1944_PNAS_30_244, Book_Murnaghan, 1995_IJT_16_1009} with the GET. Thermodynamic scaling arises in the non-Arrhenius relaxation regime as a scaling property of the fluid configurational entropy density $s_c$, normalized by its value $s_c^*$ at the onset temperature $T_A$ of glass formation so that a constant value of $TV^{\gamma_t}$ corresponds to a \textit{reduced isoentropic} fluid condition. Unfortunately, a direct experimental test of this interpretation of the origin of thermodynamic scaling in real materials is made difficult by the extreme difficulty in estimating $s_c$ experimentally.~\cite{2022_Mac_55_8699} The approximation of $s_c$ by the difference between the fluid entropy and that of the low temperature crystal or solid glass is only a rather rough approximation in polymeric materials.~\cite{2013_JCP_138_12A541}

Figure~\ref{Fig_Scaling} indicates that $\tau_{\alpha}$, $L / L_A$, $\langle u^2 \rangle$, and $G_p / k_{\mathrm{B}}T$ all exhibit thermodynamic scaling to a good approximation, which is supportive of theories of glass formation based on these quantities. The scaling exponent $\gamma_t$ is smaller for stiffer chains, consistent with the prediction based on the GET.~\cite{2013_JCP_138_234501} We find that thermodynamic scaling does not hold for $G_{\infty} / k_{\mathrm{B}} T$, $S(0)$, and $\Pi_{\mathrm{CED}}$. Thermodynamic scaling properties for some of the quantities, such as $\tau_{\alpha}$, $L / L_A$, $1 / \langle u^2 \rangle$, $G_p / k_{\mathrm{B}} T$, $\alpha_2^*$, $G_{\infty} / k_{\mathrm{B}} T$, and $S(0)$, have been discussed in our previous works,~\cite{2021_Mac_54_3247, 2022_Mac_55_8699} but these results are included in Figure~\ref{Fig_Scaling} to make a comparison with our new data, obtained under the same thermodynamic conditions.

The absence of thermodynamic scaling does not by itself preclude a relationship between the structural relaxation time and other dynamical properties and thermodynamic properties showing exceptions to thermodynamic scaling. We have shown recently that, by subjecting this type of thermodynamic property to a linear transformation, thermodynamic scaling can be recovered to a good approximation in the transformed variable.~\cite{2022_Mac_55_8699, 2021_Mac_54_3247} In particular, it was shown that $B$ can be related in this way to $\langle u^2 \rangle$ to a good approximation. This relationship, which we do not discuss here, has been further verified recently for a model of thermoset polymers having variable cross-link density and cohesive interaction strength.~\cite{2022_JCP_157_064901, 2022_Mac_55_9990} Further study is required to understand the significance of these property transformations and their impact on thermodynamic scaling. We have offered some tentative ideas regarding this phenomenon.~\cite{2021_Mac_54_3247}

\subsection{\label{Sec_PICED}Fundamental Importance of Cohesive Interactions in Polymer Glass formation}

\begin{figure*}[htb]
	\centering
	\includegraphics[angle=0,width=0.975\textwidth]{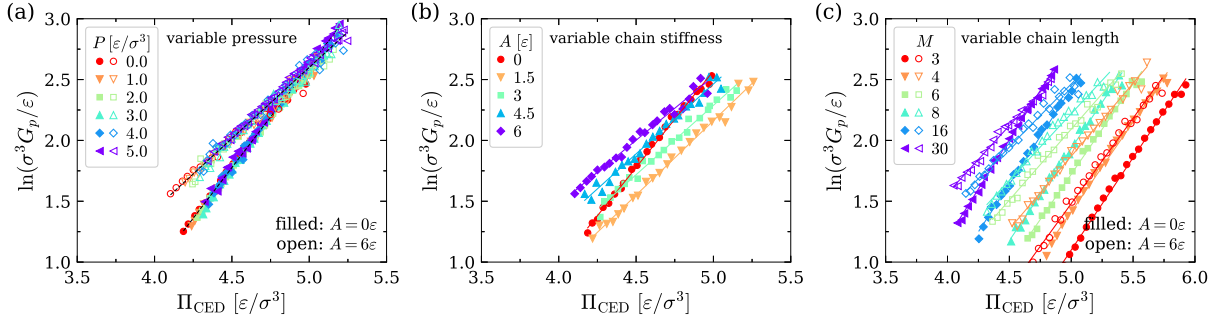}
	\caption{\label{Fig_Gp_PICED}Correlation between the glassy plateau shear modulus and the cohesive energy density. Panels (a--c) show $\ln(G_p)$ versus $\Pi_{\mathrm{CED}}$ for variable $P$, $A$, and $M$, respectively. Filled and open symbols in panels (a) and (c) correspond to the results for $A = 0 \varepsilon$ and $6 \varepsilon$, respectively. Lines are a guide to the eye. The analysis considers the data below $T_A$.}
\end{figure*}

We take an opportunity to quantitatively check Tobolsky's intuitive idea that the glassy plateau shear modulus $G_p$ should be related to the cohesive energy density $\Pi_{\mathrm{CED}}$. As shown in Figure~\ref{Fig_Gp_PICED}, $G_p$ varies exponentially with $\Pi_{\mathrm{CED}}$ as pressure, chain stiffness, and chain length are varied. The prefactor of the exponential and the argument of the exponential would appear to be system specific, however. Nonetheless, the observations of Figure~\ref{Fig_Gp_PICED} strongly support the hypothesis of Tobolsky that the glassy modulus should be dominated by the cohesive interaction strength of the material.~\cite{Book_Tobolsky} The importance of the relationship between $G_p$ and $\Pi_{\mathrm{CED}}$ has been emphasized by Leporini and coworkers,~\cite{2015_JPSB_53_1401} and we examine this type of relationship more quantitatively in the present paper.

\begin{figure*}[htb]
	\centering
	\includegraphics[angle=0,width=0.975\textwidth]{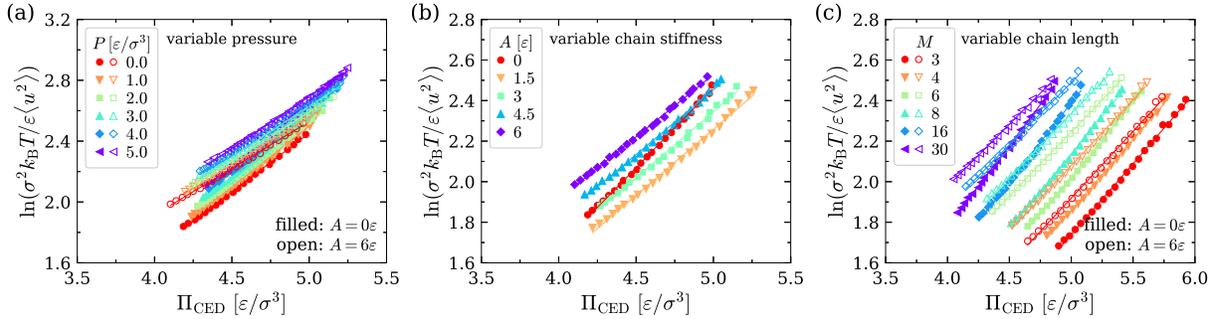}
	\caption{\label{Fig_DWF_PICED}Correlation between the Debye-Waller parameter and the cohesive energy density. Panels (a--c) show $\ln(k_{\mathrm{B}}T / \langle u^2 \rangle)$ versus $\Pi_{\mathrm{CED}}$ for variable $P$, $A$, and $M$, respectively. Filled and open symbols in panels (a) and (c) correspond to the results for $A = 0 \varepsilon$ and $6 \varepsilon$, respectively. Lines are a guide to the eye. The analysis considers the data below $T_A$.}
\end{figure*}

We note that a recent method of coarse-graining the dynamics of polymer liquids in a temperature transferable fashion, the energy renormalization method,~\cite{2019_SciAdv_5_eaav4683} relies on the sensitivity of $\langle u^2 \rangle$ to the cohesive interaction parameter $\epsilon$. The success of this method suggests to us that there might be some general relation between $\langle u^2 \rangle$ and $\Pi_{\mathrm{CED}}$, which would imply an important linkage between the dynamic and thermodynamic properties of fluids if such a relation holds generally. Given the finding in Figure~\ref{Fig_Gp_PICED} and the linear relation between $G_p$ and $k_{\mathrm{B}} T / \langle u^2 \rangle$ in eq~\ref{Eq_GpDWF}, a linear relation should be expected between $\ln (k_{\mathrm{B}} T / \langle u^2 \rangle)$ and $\Pi_{\mathrm{CED}}$, which we confirm in Figure~\ref{Fig_DWF_PICED}. We had not initially anticipated these relations between $G_p$, $\langle u^2 \rangle$, and $\Pi_{\mathrm{CED}}$, which deserve some theoretical consideration.

\subsection{\label{Sec_Quantification }Elastic Heterogeneity and Its Potential Relevance to the Nonlinear Deformation Properties of Polymer Glass Materials}

Previous simulation studies of polymeric~\cite{2013_JCP_138_12A541} and metallic~\cite{2015_JCP_142_164506} GF liquids have established that the phenomenon of dynamic heterogeneity is physically realized by clusters of particles of relatively high and low mobility having a fractal structure, whose average sizes grow upon cooling. The geometry of these clusters is rather similar to that in associating particle systems that form branched polymers at equilibrium.~\cite{2018_SoftMatter_14_1622, 2016_JCP_144_074901} The strings generally conform to linear chain equilibrium polymers~\cite{2003_JCP_119_12645, 2006_JCP_125_144907, 2007_JCP_126_194903, 2008_JPCM_20_155101, 2014_JCP_140_204509, 2018_JCP_148_104508, 2021_Mac_54_3001} in the systems studied so far, but the universality of this phenomenon remains to be established for other GF liquids, especially network-forming glass-formers, such as silica and water.

\begin{figure*}[htb]
	\centering
	\includegraphics[angle=0,width=0.975\textwidth]{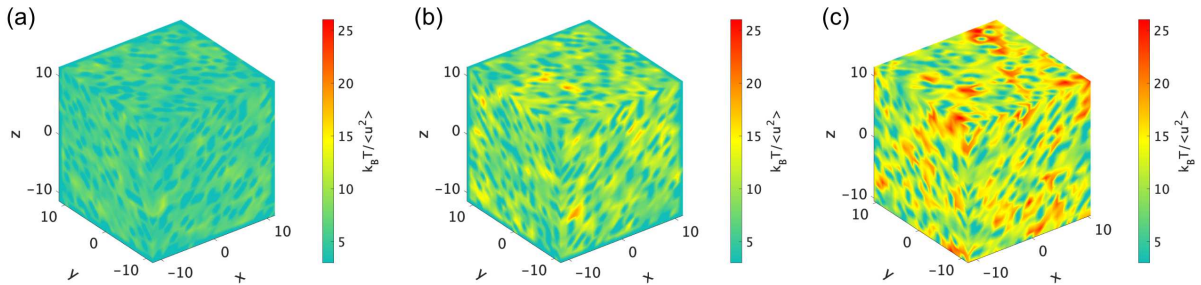}
	\caption{\label{Fig_StiffnessA}Color maps of local molecular stiffness $k_{\mathrm{B}} T / \langle u^2 \rangle$ of particles at $k_{\mathrm{B}} T / \varepsilon = 0.66$ for polymer melts having variable $A$. Panels (a--c) correspond to results for $A = 0 \varepsilon$, $3 \varepsilon$, and $6 \varepsilon$, respectively. The chain length is fixed at $M = 20$ and the pressure is fixed at $P = 0.0 \varepsilon / \sigma^3$.}
\end{figure*}

In the present work, we do not repeat this type of analysis considered previously and focus instead on measures of ``elastic heterogeneity''~\cite{2022_JCP_157_064901, 2022_Mac_55_9990} in relation to the understanding of overall elastic properties of our model GF liquids discussed above. Our analysis follows the recent works of Zheng et al.~\cite{2022_JCP_157_064901, 2022_Mac_55_9990} and Wang et al.,~\cite{2021_JCP_155_204504} which take eq~\ref{Eq_GpDWF} as a defining relationship for the local material stiffness. In particular, we consider $k_{\mathrm{B}} T/ \langle u^2 \rangle$ to be a measure of the ``local stiffness'' of the material to avoid ambiguities in defining a local shear modulus. More discussion can be found in the works of Zheng et al.,~\cite{2022_JCP_157_064901, 2022_Mac_55_9990} where the quantification of these local elastic heterogeneity fluctuations is discussed at length. Notably, nanoscale stiffness fluctuations have been directly observed in nanoprobe measurements on polymer films~\cite{2015_SoftMatter_11_1425} and metallic glass materials,~\cite{2011_PRL_106_125504, 2019_MRL_7_305} where the qualitative appearance of such fluctuations appears to be very similar to the results in Figure~\ref{Fig_StiffnessA}. One complication in comparing simulations to experiments is that the nature of these fluctuations can be expected to be significantly different near the boundary of the material than from the material interior so that simulations of materials with free boundaries are required to compare directly with measurements of nanoscale stiffness fluctuations on real glassy materials.

Based on the direct visualization of the elastic heterogeneity in our polymer models, we may gain insight into some of the quantities discussed above associated with correlation functions measured experimentally. In previous works,~\cite{2016_Mac_49_8355, 2017_Mac_50_2585, 2020_Mac_53_4796} we and others before us have shown that the distribution function of the immobile particle clusters, corresponding to regions of high local stiffness in our stiffness ``maps'', peaks at a time that is essentially equivalent to the structural relaxation time $\tau_{\alpha}$. The $\chi_{4, s}$ function peaks at nearly the same time, and this correlation function is correspondingly dominated by the immobile particles and thus provides information about this particular type of dynamic heterogeneity. At low $T$, $\tau_{\alpha}$ becomes large and the immobile particle clusters can then be viewed as elastic regions, hence the term ``elastic heterogeneity''.~\cite{2022_JCP_157_064901, 2022_Mac_55_9990} Correspondingly, the lifetime of the mobile particle clusters is a much shorter time generally in GF liquids at low $T$ and this timescale correlates strongly with the peak time of the non-Gaussian parameter and the peak time at which the strings are defined.~\cite{2013_JCP_138_12A541, 2016_Mac_49_8355, 2017_Mac_50_2585} Moreover, the diffusion coefficient in atomic fluids correlates strongly with the lifetime of the mobile particle clusters~\cite{2013_JCP_138_12A541, 2015_JCP_142_164506} so that the non-Gaussian parameter apparently provides information about the persistence time of the mobile particle clusters, which ``live'' in the interstitial region between the locally stiff regions indicated in Figure~\ref{Fig_StiffnessA} above. Decoupling is thus just a relation between the average lifetimes of the mobile and immobile particle regions or equivalently the ``soft'' and ``stiff'' regions of the material.

It is evident from the visualization of elastic heterogeneity that GF materials are highly heterogeneous elastically, despite their normally relatively high uniformity from a density standpoint. This important point was made earlier by Riggleman et al.,~\cite{2010_SoftMatter_6_292} who first constructed this type of stiffness map in GF liquids. This heterogeneity has many implications that could be quantified to give insight into basic and general properties of GF liquids. The existence of a disparity in the well-defined mobility regions in the fluid persisting over a long timescale is sufficient to give rise to the breakdown of the Stokes-Einstein relation, a phenomenon that can be rationalized from the known properties of a solution of droplets whose viscosity is different from the surrounding medium.~\cite{1998_JNCS_235_237_137} This effective medium hydrodynamic calculation predicts that $\zeta = 0.4$ when the immobile regions have a very low mobility and lower values of $\zeta$ are predicted by this model when the dynamic contrast between the immobile particle clusters and the surrounding medium is lower. The emergence of a growing stiffness in cooled liquids is clearly associated with the growth of large clusters of the finite immobile particles, and we may expect the measure of this type of dynamic heterogeneity, the peak time of $\chi_{4,s}$ to track the structural relaxation time, and its magnitude to grow with cooling. Moreover, the magnitude of $\chi_{4,s}$ should be highly sensitive to finite size effects associated with this rigidity percolation phenomenon. A high sensitivity of $\chi_{4,s}$ to finite size effects has been seen in a model GF liquids,~\cite{2009_PNAS_106_3675} as naturally expected from the current qualitative discussion. Correspondingly, we would expect that the non-Gaussian parameter $\alpha_2(t)$, which strongly correlates with the lifetime of mobile particles, along with the string length, should be relatively insensitive to finite size effects. Given the direct relation between the structural relaxation time and the string length discussed above, we may expect by extension that the $\alpha$-relaxation time should have little or no finite size dependence. A weak dependence of the structural relaxation time to finite size has indeed been observed in simulations of a model GF liquid.~\cite{2009_PNAS_106_3675} This is another topic that deserves to be revisited because of its evident theoretical and practical importance in relation to the dynamics of GF liquids from a more fundamental perspective.

We may understand the development of the ubiquitous stretched exponential stress relaxation of GF liquids and the occurrence of the $\alpha$-$\beta$ bifurcation of the relaxation times governing an initial fast and slow $\alpha$-relaxation process in highly cooled liquids based on standard models of structural relaxation in polymer solutions and thermodynamic models of self-assembly into dynamic polydisperse polymer clusters upon cooling,~\cite{1991_Mac_24_3163, 2008_JCP_129_094901} as directly observed in simulation studies.~\cite{2013_JCP_138_12A541, 2019_JCP_151_184503, 2015_JCP_142_164506, 2021_EPJE_44_56} It is also straightforward to understand that instabilities might initiate in the soft regions when the materials are deformed, as shown in recent simulations where the stiffness map provided insight into incipient shear banding in a model metallic glass material.~\cite{2021_JCP_155_204504}

\begin{figure*}[htb]
	\centering
	\includegraphics[angle=0,width=0.975\textwidth]{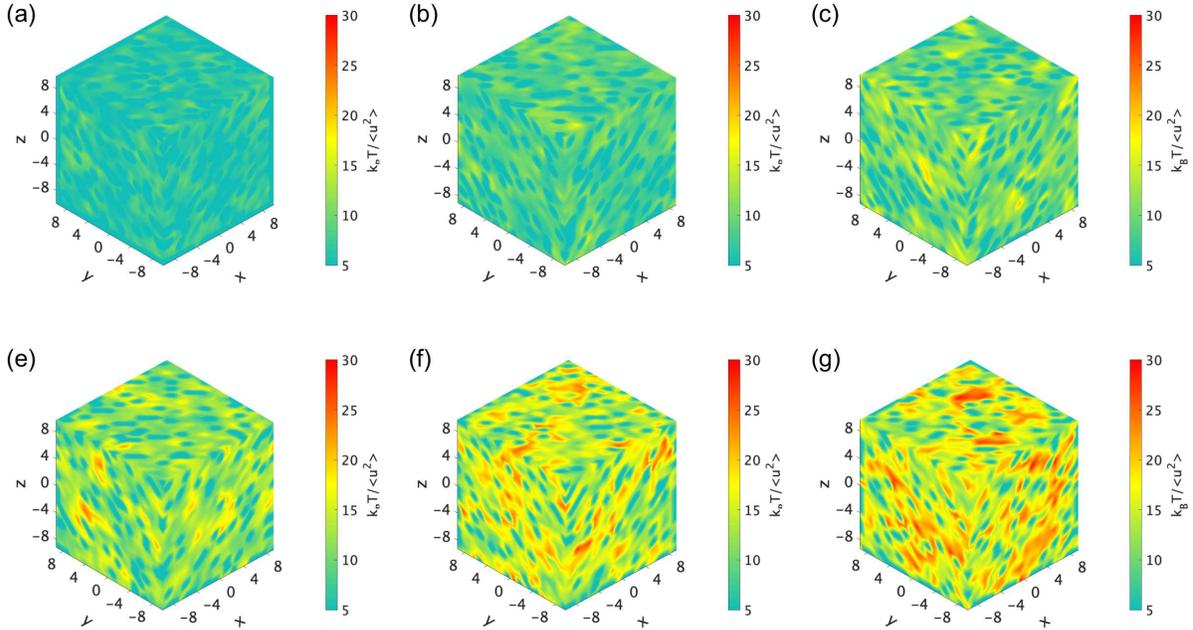}
	\caption{\label{Fig_StiffnessP}Color maps of local molecular stiffness $k_{\mathrm{B}} T / \langle u^2 \rangle$ of particles at $k_{\mathrm{B}} T / \varepsilon = 0.61$ for polymer melts having variable $P$. Panels (a--g) correspond to results for $P = 0.0 \varepsilon / \sigma^3$, $1.0 \varepsilon / \sigma^3$, $2.0 \varepsilon / \sigma^3$, $3.0 \varepsilon / \sigma^3$, $4.0 \varepsilon / \sigma^3$, and $5.0 \varepsilon / \sigma^3$, respectively. The chain length is fixed at $M = 20$ and the chain rigidity is fixed at $A = 0 \varepsilon$.}
\end{figure*}

We further expect that large stiffness fluctuations in some polymeric GF liquids have a particular significance in understanding plastic deformation of this broad class of materials. It has long been observed that subjecting polymer materials in the glass state to a large deformation leads to strain stiffening~\cite{2008_JPSB_46_2475} and that this deformation is often reversible when the material is allowed to relax above $T_{\mathrm{g}}$. Treloar~\cite{1944_RCT_17_813} and Vincent~\cite{1960_Polymer_1_7} argued that there are strong cohesive regions in the glass state which play the role of effective physical cross-links. Haward and Thackery~\cite{1968_PRSLA_302_453, 1993_Mac_26_1860} formulated a theory of the large deformation of ``plastic'' polymer materials based on classical rubber elasticity with the cross-linking density taken to be phenomenological and distinct from the chemical cross-linking density. This model has notably found applications in both semi-crystalline~\cite{1995_JPSB_33_1481, 1983_JMS_18_903, 1993_Mac_26_1860} and GF polymer materials.~\cite{2003_Polymer_44_2493, 2004_JPSB_42_2041, 2000_JR_44_1263, 2008_JPSB_46_2475} We suggest that the phenomenology of these materials, and some of the models used extensively to rationalize their mechanical properties under large deformation, can naturally be explained by the existence of the emergence of stiff regions in the GF liquids which would play the role of effective physical cross-links between the chains, just as local crystallites in semi-crystallline polymers have long been suggested to rationalize the effective cross-links in these materials. This physical interpretation of the large deformation properties of both amorphous and crystalline polymer materials in their solid state would naturally imply that the linkages would progressively ``melt'' upon heating, as evidenced by a vanishing yield strength near their $T_{\mathrm{g}}$~\cite{2004_JPS_42_2050, 2006_APL_88_221109} or melting temperatures, respectively. This is exactly the phenomenology commonly observed in amorphous polymer materials in their glass state.~\cite{2003_Polymer_44_2493, 2004_JPSB_42_2041, 2000_JR_44_1263, 2008_JPSB_46_2475} A simple model~\cite{2008_JPSB_46_2475, 1972_Polymer_13_174} has been introduced in which dynamic associations are responsible for the cohesive linkages, a concept hypothesized long ago by Treloar~\cite{1944_RCT_17_813} and Vincent.~\cite{1960_Polymer_1_7} Lin et al.~\cite{2010_SoftMatter_6_3548} proposed a model having a similar spirit to study polymer composites in which the particles are assumed to create physical cross-links between the chains. This work also considered the effect of deformation on the association and rate effects associated with stress relaxation that can be expected to be operative in deformed glassy materials. It is evident from these models that both enthalpic and entropic effects associated with the deformation of the polymer chains in the networks, as in rubber elasticity, along with relaxation effects, should play a role in the resulting viscoelastic response of solid polymer materials under large deformation. Enthalpic contributions to the strain stiffening in amorphous polymer materials deformed beyond their yield point have been emphasized in simulations by Hoy and Robbins~\cite{2007_PRL_99_117801} in their critique of the Haward-Thackery theory~\cite{1968_PRSLA_302_453, 1993_Mac_26_1860} and its extensions emphasizing entropically-dominated elasticity. Recent measurements clearly indicate that entropic elasticity is at least a contributing factor to the large deformation response of plastic polymer materials.~\cite{2008_JPSB_46_2475} Neutron scattering~\cite{2008_Mac_41_860} and nuclear magnetic resonance measurements~\cite{2006_JMPS_54_589} have provided evidence of chain deformation consistent with a polymer network model of reversible elastic deformation in a highly deformed polymer melt when the material is annealed above $T_{\mathrm{g}}$ to facilitate the network relaxation process. Wang et al.~\cite{2014_JCP_141_094905, 2012_Mac_45_6719} have introduced an interesting conceptual model of the plastic deformation of amorphous polymers in their glass state and the brittle ductile transition in tensile deformation in this important class of structural materials that assumes a ``dual network'' in these materials composed of primary polymer associations due to local short-range molecular interactions, and a ``vitrified entanglement network'' acting to confine the chains at larger scales deriving from the topological interactions associated with the uncrossability of the chains locked into the physical network within the polymer glass. Our simulation findings suggest a similar physical picture to that suggested by Wang et al.,~\cite{2014_JCP_141_094905, 2012_Mac_45_6719} except that local stiffness fluctuations in the glass state serve as the physical cross-linking agency rather than the chain hairpins suggested in the works of Wang et al.~\cite{2014_JCP_141_094905, 2012_Mac_45_6719} As a specific and novel example of the value of this type of visualization of elastic heterogeneity fluctuations to gain insight into nonlinear mechanical properties of polymers, we start from a general observation that amorphous fragile polymer glasses such as polystyrene and poly(methyl methacrylate) become highly ductile when these materials are subjected to high applied pressures,~\cite{Chapter12_PolymerGlasses} a well-known physical effect that has not been explained up to the present. Figure~\ref{Fig_StiffnessP} shows the effect of progressively increasing the pressure on stiffness fluctuations in our coarse-grained melt at a fixed $T$. We see that the intensity of the elastic constant fluctuations strongly increases with increasing pressure, which can naturally be expected to increase the density of physical cross-link regions corresponding to regions stiff enough to act as cross-links in the glass material. We then have a clear possible physical rationale for how increasing pressure should alter the large deformation of amorphous polymer glass materials, as observed experimentally.~\cite{1964_Nature_202_381, 1976_JAPS_20_1853} In a similar fashion, we could readily utilize the stiffness field to understand aging effects in polymer glasses, the effects of nanoparticle and solvent additive, polymer deformation by applied mechanical deformation, electric or magnetic field, etc.

We mention that the above perspective of GF polymer materials also provides an understanding of a recent report~\cite{2021_Nature_596_372} in which such materials exhibit ``rubbery'' entangled-like surface behavior even in polymers composed of short sub-entangled chains. We may expect the physical cross-links arising from stiffness fluctuations to account for the rubbery consistency of these materials, where we may also expect a gradient in these elastic fluctuations and thus stiffness near the free boundary of the material. Numerous studies have shown that the molecular stiffness, $k_{\mathrm{B}} T / \langle u^2 \rangle$, thus the local stiffness, is decreased near the material boundary over a temperature dependent scale in both GF and crystalline materials.~\cite{2013_SoftMatter_9_241, 2013_SoftMatter_9_1266, 2018_SoftMatter_14_8814, 2020_JCP_153_124508, 2021_EPJE_44_33, 2022_JCP_157_094901} This should lead to a nontrival deformation behavior near the free boundary of glassy polymer materials under large deformation conditions that naturally arise in the contexts of tribology and adhesion so that our capacity to understand and control elastic fluctuations (e.g., tuning fragility with additives or changing polymer microstructure) and to correspondingly engineer materials based on this knowledge to optimize effects deriving from them should have many applications. This statement applies also to the engineering of plastic polymer materials for the diverse range of structural applications in which they are utilized.

We expect that future studies based on this type of mapping of elastic heterogeneity should also provide insights into other basic aspects of GF liquids, such as physical aging, stress and field-induced changes in local molecular mobility, changes in the material properties with polymeric and molecular additives, and changes in gas permeability with molecular structure and complex phenomena, where we may likewise expect elastic heterogeneity at a nanoscale to play an important role. We leave such studies to future work.

\section{Summary}

The theoretical development of models of glass formation has been an ongoing saga based largely on phenomenological correlations between thermodynamic properties, such as density, compressibility, entropy, shear and bulk moduli, and enthalpy of liquids, and dynamic properties, such as viscosity, diffusion coefficients, and structural relaxation times. The persistence of the many models developed over the years can be attributed to the success of each of these models in rationalizing observations on real materials. In the present work, we have extended our previous efforts to achieve a more unified treatment of the dynamics of glass-forming liquids through an inclusion of how the shear and bulk moduli relate to other thermodynamic properties and basic dynamical properties of coarse-grained polymer melts having a range of chain stiffnesses and cohesive interaction strengths for a wide range of temperatures and pressures to establish relationships of generality.

Most theories of the dynamics of glass-forming liquids are predicated on the assumption of activated transport, as in transition state theory. The large change in the relaxation time, viscosity, and diffusion coefficient of glass-forming liquids is thus attributed to the temperature dependence of activation energy. The differences in these models relate models of the activation energy in terms of the collective motion and associated changes of configurational entropy to emergent rigidity in the cooled liquid modeled in terms of the bulk or shear modulus or a local measure of rigidity in the material defined in terms of the Debye-Waller parameter. We find that all these models are closely interrelated so that each of these models is potentially ``valid''.

Specifically, we have computed basic dynamical properties for our family of polymer models, such as the shear stress relaxation function $G(t)$ and the self-intermediate scattering function $F_s(q, t)$ to characterize relaxation in these materials. We have also calculated the non-Gaussian parameter $\alpha_2(t)$ and the four-point density correlation function $\chi_{4, s}(t)$ to quantify the dynamic heterogeneity of these materials associated with mobile and immobile particle clusters, respectively. From these current standard dynamic heterogeneity measures, we obtain basic relaxation times related to the dynamics of the material and the ``stiffness measures'', such as the glassy plateau shear modulus $G_p$ and the ``non-ergodicity parameter'' $f_{s, q^*}$ from the transient plateauing of the relaxation functions following the fast $\beta$-relaxation on a ps timescale of both of these relaxation functions. The mean square displacement $\langle u^2 \rangle$ at the same ``caging'' timescale, a quantity directly related to $f_{s, q^*}$, is also emphasized since this quantity can be measured through many experimental methods and has frequently been interpreted in recent modeling of glass-forming liquids as a measure of material stiffness that has been suggested to be of relevance for understanding the characteristically strong temperature dependence of the structural relaxation time of glass-forming liquids. The glassy shear modulus $G_p$ is another obvious measure of material stiffness that has likewise been emphasized as being crucial for understanding the temperature dependence of the relaxation time of glass-forming liquids. To enable the comparison of measures of material stiffness with measures of collective motion that have been shown to be highly correlated with changes in the relaxation of glass-forming liquids, we fit our relaxation time data for all the polymer models and thermodynamic conditions considered to the string model of glass formation which provides quantitative information about how the scale of collective motion (i.e., string length, $L$) changes with temperature. We further verify that the relaxation times obtained from $G(t)$ and $F_s(q, t)$ are consistent in the glassy dynamics regime of our polymer melt simulations, justifying the identification of $\tau_{\alpha}$ as the ``structural relaxation time''. 

In addition to showing that $\tau_{\alpha}$ can be quantitatively described by models emphasizing both stiffness measures and the scale of collective motion, we have shown that the scale of collective motion and the rigidity measures are correspondingly interrelated, as required by mathematical consistency between these models of the dynamics of glass-forming liquids. The emergent elasticity of cooled liquids is evidently matched by emergent collective motion. Moreover, we have also shown that the macroscopic stiffness parameters, $G_p$ and $f_{s, q^*}$, and for completeness, the bulk modulus $B$ of the material, can all be expressed \textit{quantitatively} in terms of molecular scale stiffness parameter, $k_{\mathrm{B}} T / \langle u^2 \rangle$, so that the short time dynamics of the fast $\beta$-relaxation on a ps timescale provides an understanding of both the long-time structural relaxation time $\tau_{\alpha}$ and the macroscopic rigidity of the material. This interrelation between macroscale and molecular scale measures of rigidity led us to consider ``stiffness maps'' of the materials that allow us to visualize the long-lived elastic heterogeneity of our cooled polymer melts when the lifetimes of these regions become very long, as it naturally does upon approaching the glass transition temperature. This construct allows us to clearly visualize the fluctuations in dynamic heterogeneity that is characteristic of glass formation to better understand the relaxation times and other properties measured for these liquids through simulation or experiment. We expect that this information should be very useful for better comprehending many aspects of the dynamics of glass-forming materials and in the design of new materials.

In the course of our systematic study of the elastic properties of coarse-grained polymer melts, we have discovered unexpected relations between $G_p$ and a particularly important thermodynamic property, the cohesive energy density $\Pi_{\mathrm{CED}}$, for all the materials and conditions that we investigate. This observation underscores the importance of attractive intermolecular interactions in both the dynamics and elastic properties of condensed materials.~\cite{2021_Mac_54_3247} We have also observed an unanticipated relation between the peak height of the mysterious dynamic heterogeneity measure given by $\alpha_2(t)$ and $G_p$, which points to a need to better understand both of these properties of glass-forming liquids.

We have examined the extent to which all the properties that we simulate conform to the thermodynamic scaling property observed in the temperature and density dependence of the relaxation times of our material and in many real glass-forming liquids. This analysis provides support for the hypothesis of Tobolsky~\cite{Book_Tobolsky} that the glassy modulus should be dominated by cohesive interactions between the chains while the infinite frequency modulus should reflect the chain bonding interactions and thus should have very different values and even a qualitatively different temperature dependence. We also gain some insights into why some thermodynamic and dynamic properties exhibit thermodynamic scaling, while others do not, information that is germane to the development of an acceptable theory of glass formation that reproduces the thermodynamic scaling property observed in most glass-forming materials.

\begin{acknowledgement}
W.-S.X. acknowledges the support from the National Natural Science Foundation of China (Nos. 22222307 and 21973089). X.X. acknowledges the support from the National Natural Science Foundation of China (Nos. 21873092 and 21790341). The authors are grateful to Prof. Wenjie Xia for help with the visualization of stiffness maps. W.-S.X. and X.X. are grateful to Dr. Teng Lu for help with the numerical calculations conducted on the ORISE Supercomputer. This research used resources of the Network and Computing Center at Changchun Institute of Applied Chemistry, Chinese Academy of Sciences.
\end{acknowledgement}


\bibliography{refs}

\end{document}